\documentclass[12pt]{iopart}
\usepackage{url}
\usepackage{graphicx}
\usepackage{soul}
\usepackage{cite}
\usepackage{hyperref}
\usepackage{xcolor}
\usepackage{caption}
\usepackage{hhline}
\usepackage{setspace}

\begin{document}

\title[Multi-Year-to-Decadal Machine Learning Model-Analog Prediction]{Multi-Year-to-Decadal Temperature Prediction using a Machine Learning Model-Analog Framework}
\author{M. A. Fernandez$^{1,2}$ and Elizabeth A. Barnes$^{1,2}$}
\address{$^1$Department of Atmospheric Science, Colorado State University, Fort Collins, CO, USA}
\address{$^2$Faculty of Computing and Data Sciences, Boston University, Boston, MA, USA}
\ead{mafern@colostate.edu}
\begin{abstract}
Multi-year-to-decadal climate predictions are a key tool in understanding the range of potential regional climate futures.
Here, we present a framework that combines machine learning and analog forecasting for regional predictions on these timescales.
A neural network is used to learn a mask of weights that highlights important global precursors to the evolution of a specific prediction target (region, variable, and lead time).
A library of mask-weighted model states, or potential analogs, are then compared to a mask-weighted observational state.
The known future of the best matching potential analogs serve as the prediction for the future of the observational state.
We predict 2-meter temperature using the Berkeley Earth Surface Temperature dataset for observations, with a multi-model potential analog library of CMIP6 simulations.
Using a 30-year climatology reference, we compare our analog method to two other analog prediction methods and the CMIP6 library using the continuous ranked probability score to assess the quality of predicted distributions, and mean squared error to assess the quality of predicted ensemble means.
For nearly all cases explored, our analog method produces skillful predictions.
We find higher distribution skill over the CMIP6 library in all cases, which is due to improved prediction of the distribution spread.
We find overall higher skill than other analog methods in the predicted distributions and ensemble means.
Finally, we find broadly similar skill to an ensemble of bias-corrected initialized Earth system models.
Benefits of our analog method include low computational cost, ensemble size flexibility, and interpretablility.
\end{abstract}
\vspace{2pc}
\section{Introduction}\label{sec:intro}

Skillful predictions of regional climate from years to decades are necessary for understanding both the future climatological mean and potential extremes \cite{wmo, IPCC_WG1}.
These predictions are therefore vital for climate services, including mitigation and adaptation planning on regional scales \cite{Khasnis2005, Solaraju2022, Dunstone2022, IPCC_WG2, IPCC_WG3}.

There are many methods aimed at predicting future weather and climate on multi-year to decadal timescales.
Many of these make use of the large collection of available Earth System Models (ESMs) that have been developed at research centers around the world.
Initialized ESMs (IESMs) use observations to initialize physics-based dynamical models, then run these simulations, often out to ten years \cite{boer2016, meehl2021}.
However, due to model biases and unresolved physics, IESMs often require some form of bias-correction \cite{meehl2021, meehl2022}.
IESMs can also be costly to run, making large ensembles challenging to create and potentially reducing their effectiveness for the prediction of extremes.

Work in the last several years \cite{Befort2020, Mahmood2021, Befort2022, DeLuca2023, Donat2024} has used both observations and IESMs to constrain climate projections of the next 10 to 20 years.
These studies match the projections to either observations or IESMs over a common period for a specified region, keeping only the best matching members to create a constrained sub-ensemble.
These constrained sub-ensembles have been shown to have better skill than the full ensemble for up to $20$ years.

An alternaitve initialization method that has a long history is analog forecasting \cite{Lorenz1969}.
In analog forecasting, an initial observation-based or model-based state is matched to a library of potential analogs.
The known future of the potential analogs that best match the initial state serves as the prediction.
Analog forecasting is time-agnostic, e.g., the prediction for $2026$ does not need to come from model output for $2026$.
Early use of analog forecasting was for weather prediction, using historical observations as the analogs for the current state of the system \cite{Bergen1982}.
There have been several recent studies exploring the use of ESMs in an analog forecasting framework, termed model-analogs \cite{ding2018, Rader2023, ding2023, toride2024, acosta2025}.
The underlying premise of model-analogs is that there already exists a large catalog of ESMs that can be used to match to observations, and from which analog predictions can be made.

Here, we build on work from \cite{Rader2023}, which used machine learning to identify the most important precursor features for matching in a model-analog framework.
We use an updated machine learning architecture, new matching criteria, an expanded dataset that includes members from $29$ Coupled Model Intercomparison Project Phase 6 (CMIP6; \cite{eyring2016}) models, explore more regions and longer lead times, evaluate our method against IESMs more comprehensively, include a probabilistic metric, and focus on observations as truth.
We make predictions for lead times from one to ten years in multiple regions.
We evaluate these predictions against alternative analog methods, the CMIP6 catalog, as well as bias-corrected IESMs.

\section{Prediction Framework}\label{sec:framework}

\begin{figure}[h!]
    \noindent\includegraphics[width=\textwidth]{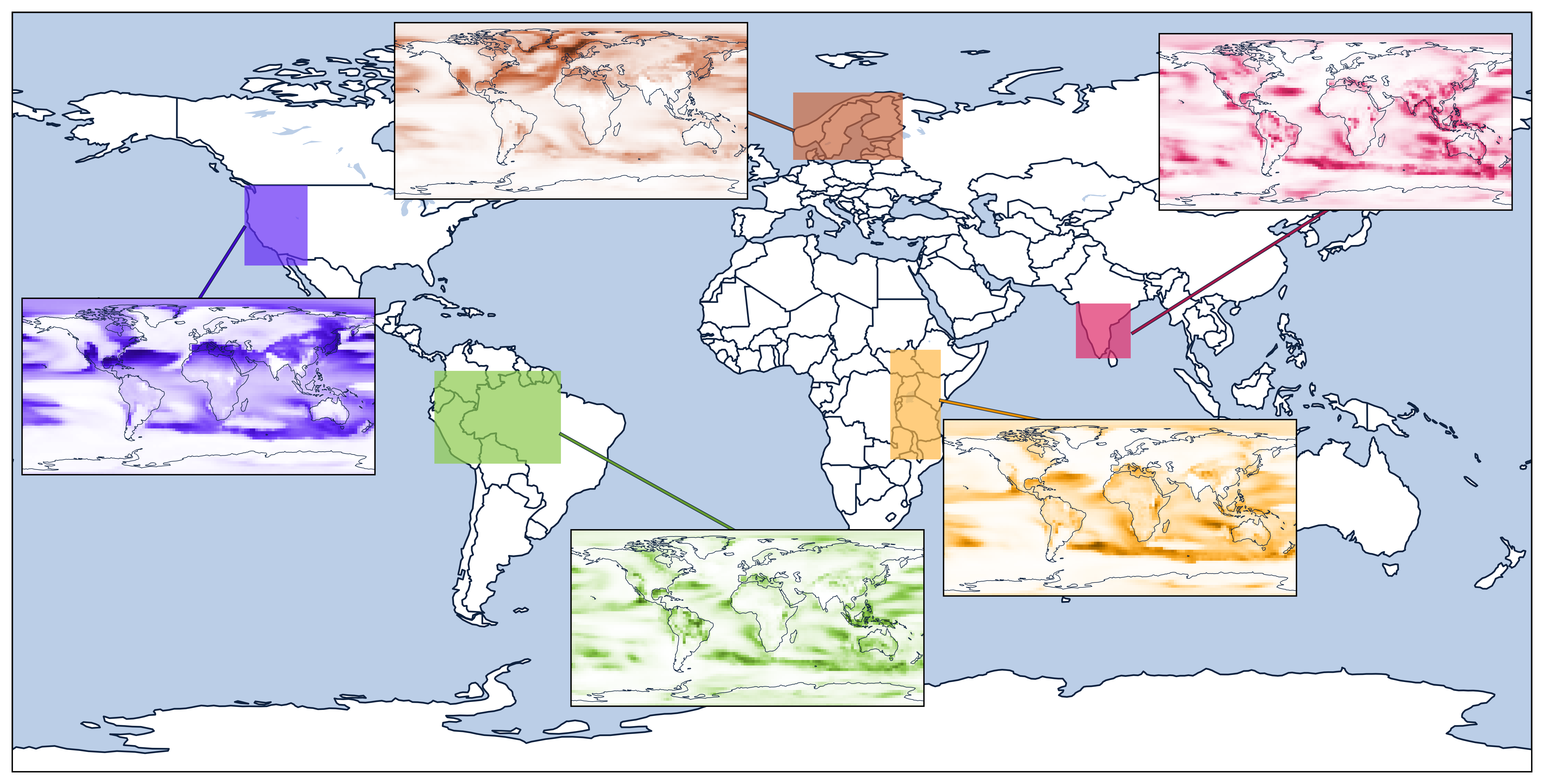}
    \caption{Example learned masks for annual mean 2-meter temperature for the five regions highlighted in the main map: the western United States, the Amazon, northern Europe, the great lakes of Africa, and southern India.
    Inset maps show examples of learned masks where darker colors indicate more importance and white indicates low importance (scale is normalized to sum to 1).
    Each region has a unique mask highlighting global locations that are important markers for the evolution of that region, on that time scale.}
    \label{fig:global_masks}
\end{figure}

Our aim is to capitalize on existing Earth system model (ESM) simulations for prediction on multi-year-to-decadal timescales.
We do this using an analog forecasting framework.
The library of analogs is composed of model simulations from the CMIP6 suite (Section \ref{sec:data}), and the method for choosing analogs emphasizes important precursor features which are determined using machine learning (Section \ref{sec:mask}) for a specific prediction task (target region, lead time, and variable).
We match on (Section \ref{sec:selection}) and predict annual mean 2-meter temperature for five regions across the globe, selected for geographic diversity.
Example learned masks, described in Section \ref{sec:mask}, are shown in Figure \ref{fig:global_masks}.

\subsection{Data}\label{sec:data}

Our dataset includes 2-meter temperature from $29$ ESMs from the CMIP6 suite of simulations, detailed in Table \ref{table:cmip6}.
Simulations are obtained from the Earth System Grid Federation (ESGF; \cite{esgf}), and at minimum include the historical period from $1850$-$2014$.
A subset contain the Shared Socioeconomic Pathways (SSP) projections corresponding to SSP-3.7.0, which runs from $2015$-$2100$ \cite{scenario_mip}.
Note, a few of these simulations do not reach $2100$, namely CAMS-CSM1 (reaches $2099$) and NorESM2-LM (reaches $2054$).
In total, there are $284$ simulations that include both the historical period and SSP projections, with an additional $181$ simulations that include only the historical period.
Observations of surface temperature are from the Berkeley Earth Surface Temperature (BEST) dataset \cite{best}.

\begin{table}[h!]
    \caption{CMIP6 dataset: total number of historical members for each model, as well as number (in parentheses) that include their associated SSP-3.7.0 projections.}\label{table:cmip6}
    \centering
    \small
    \begin{tabular}{l r | l r | l r}
        \hline
        model & members & model & members & model & members \\
        \hline
        ACCESS-CM2    & 10 (5)  & ACCESS-ESM1-5 & 40 (40) & AWI-CM-1-1  & 5 (5)   \\
        AWI-ESM-1-1   & 1  (0)  & BCC-CSM2      & 3  (1)  & BCC-ESM1    & 3  (0)  \\
        CAMS-CSM1     & 3  (2)  & CESM2         & 10 (3)  & CESM2-WACCM & 3  (3)  \\
        CMCC-CM2-HR4  & 1  (0)  & CMCC-CM2-SR5  & 11 (1)  & CMCC-ESM2   & 1  (1)  \\
        CNRM-CM6-1    & 24 (6)  & CanESM5       & 60 (45) & CanESM5-1   & 72 (20) \\
        CanESM5-CanOE & 3  (3)  & GISS-E2-1-G   & 47 (23) & GISS-E2-2-G & 11 (5)  \\
        MIROC6        & 50 (50) & MIROC-ES2H    & 3  (0)  & MIROC-ES2L  & 30 (10) \\
        MPI-ESM1-2-HR & 10 (10) & MPI-ESM1-2-LR & 30 (30) & MRI-ESM2-0  & 11 (5)  \\
        NorESM2-LM    & 3  (2)  & NorESM2-MM    & 2  (1)  & TaiESM      & 2  (1)  \\
        UKESM1-0-LL   & 17 (11) & UKESM1-1-LL   & 1  (1)  &             &         \\
        \hline
    \end{tabular}
\end{table}

Data is resampled to annual averages, leaving over $100,000$ states in the analog library, and regridded to the coarsest native resolution among the models, $2.77$ (latitude) by $2.81$ (longitude) degrees, which translates to $65$ latitudes by $128$ longitudes.
Higher resolutions ($1.5$ and $2$ degrees) were tested for five year predictions of the western United States, but did not significantly affect performance.
Data is normalized by baselining on a $30$ year ($1961-1990$) mean for each member, which allows models with consistent biases to be useful members of the analog library.

To compare our predictions with an established decadal prediction method, we obtain initialized Earth system models (IESMs) run as part of the Decadal Climate Prediction Project \cite[DCPP]{boer2016}.
We only use IESMs that have predictions covering at least $1961-2018$, include lead time $1-10$ year predictions, are from models that also appear in our CMIP6 analog library, and are available via ESGF at the time of this analysis.
This leaves four models and $70$ total members, as shown in Table \ref{table:iesms}.
There are many methods for bias-correcting IESMs \cite{meehl2022}.
Here, we bias-correct the IESMs using a lead time and initialization time dependent bias for each member.
The bias is the mean error of each ensemble member for the $15$ year period preceding and including the initialization year.
For example, all IESM predictions initialized on $1984$ have a bias-correction that is the mean error between that IESM member and the observations over the time period $1970-1984$.
IESMs are then processed using the same steps as the analog library (annual means, resolution, baselining).

\begin{table}[h!]
    \caption{
        Initialized Earth System Models (IESMs).
        Models use either full field or anomaly initialization each year and are run out to ten years \cite{boer2016, meehl2021}.
        The initialized components are those that are brought into agreement with observations.
    }\label{table:iesms}
    \centering
    \small
    \begin{tabular}{l c r}
        \hline
        model & members & initialized components \\
        \hline
        CMCC-CM2-SR5 \cite{CMCC_DCPP, meehl2021}            & 10 & (full field) atmosphere, ocean, land, sea-ice \\
        CanESM5 \cite{CanESM5_DCPP, Sospedra-Alfonso_2021}  & 40 & (full field) atmosphere, ocean, sea-ice \\ 
        MIROC6 \cite{MIROC6_DCPP, Kataoka_2020}             & 10 & (full field) atmosphere, sea-ice (anomaly) ocean \\
        MPI-ESM1-2-HR \cite{MPI_DCPP, meehl2021}            & 10 & (full field) atmosphere (anomaly) ocean, sea-ice  \\
        \hline
    \end{tabular}
\end{table}

\subsection{Mask of Weights}\label{sec:mask}

In order to identify important precursor features on which to match an initial state with the library of analogs, we follow \cite{Rader2023} and turn to machine learning.
Broadly, a lightweight neural network with a global mask layer is trained on annual mean global states and their corresponding future annual mean regional states such that large differences in the global states correspond to large differences in the future regional states.

Preprocessing data for the machine learning task begins by splitting the CMIP6 data into three sets: an analog library (maximum of five members per model, all members used if model has five members or fewer), a training library (maximum of three members per model, not all models represented), and a validation library (maximum of two members per model, not all models represented).
The analog library is split off from the full set of CMIP6 simulations first, followed by the training library, and finally the validation library.
This ensures that the analog library contains at least one member from every model (there are no repeated members between the three libraries).
The analog library is also used with the validation library for early stopping.
Each of these libraries is processed into an input set, which is global, and a target set, which contains only the prediction region.
Both the input and target data are normalized by their mean and standard deviation, and shifted by the prediction lead time.

\begin{figure}[h!]
    \noindent\includegraphics[width=\textwidth]{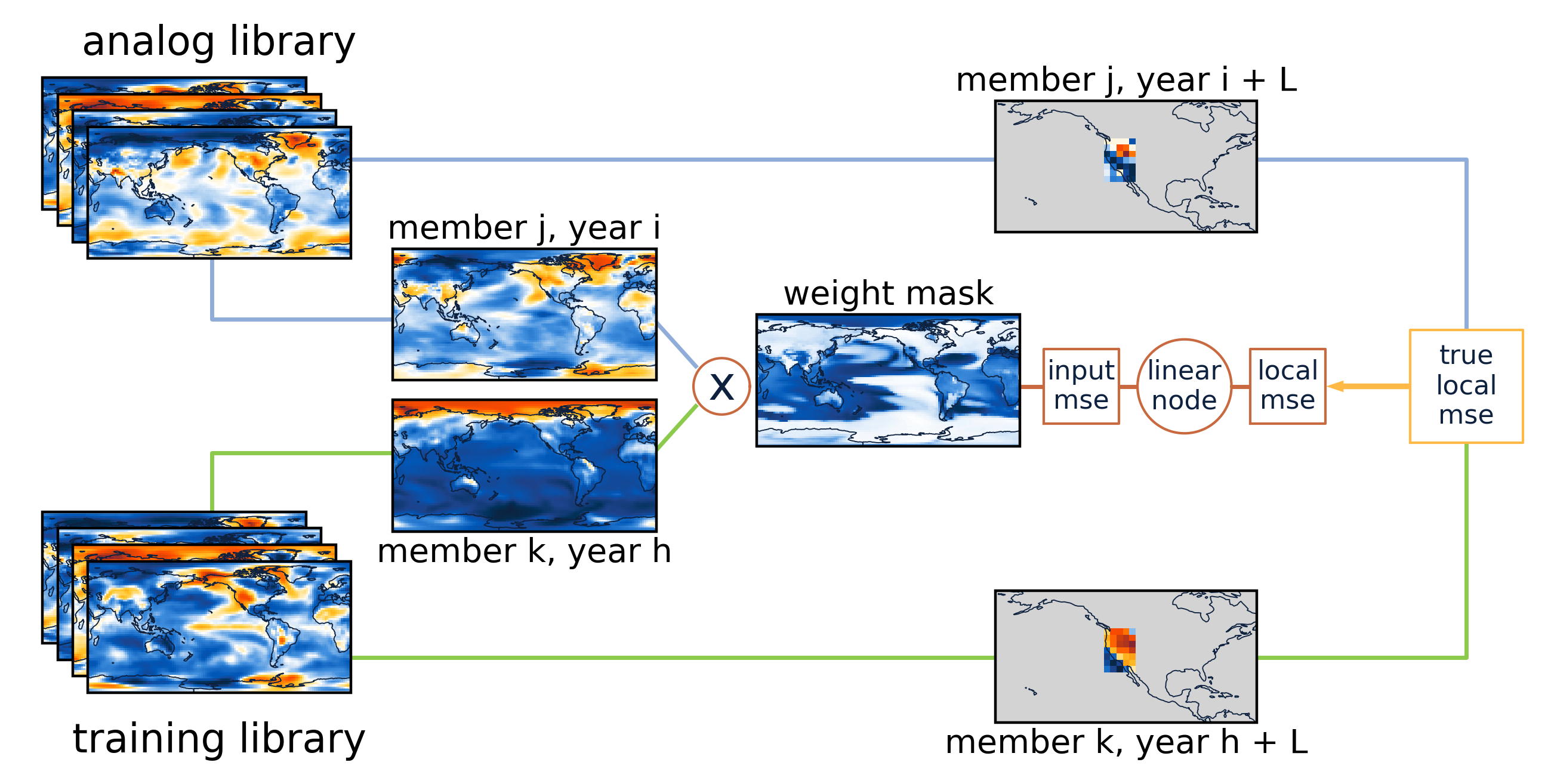}
    \caption{Schematic illustrating the machine learning component of the prediction framework.
    The learned mask identifies global precursors for predicting a specific variable for a given lead time (L years in the figure) for a specific region (western United States in the figure).}
    \label{fig:framework}
\end{figure}

Figure \ref{fig:framework} illustrates the machine learning steps after the data has been preprocessed.
Two global states, one from the analog library (e.g., member j, year i) and one from the training library (e.g., member k, year h), are selected.
The states are multiplied by a weight mask that is initialized as all ones and then updated through training.
The mean-squared error (MSE) between these two states is then passed through a single linear node which predicts the MSE of the target region at some lead time L years in the future (MSE between the target region for analog member j in year i+L and training member k in year h+L).

The mask of weights is updated through backpropagation, where the loss is the error between the predicted and true future target MSE and the learning rate is $0.001$.
We employ early stopping (patience of $50$ epochs and minimum change of $0.005$) using the MSE computed between states selected from the analog library and validation library. 
For training, the batch size is set to $64$ and the activation function for the mask is relu.

In addition to the model-learned mask, we train a second mask using transfer learning for comparison \cite{pan2010, ghani2024}.
We update the learned masks by retraining on both models (training and validation library) and a subset of the BEST observations (analog library) covering the period $1956-1998$.
The use of observations to finetune the model-learned masks is intended to correct biases that may exist in the models.
Given the interpretable nature of our learned mask method, we evaluate changes between the model-learned and transfer learned masks.
To retain information learned in the model training, the linear node (see Figure \ref{fig:framework}) that predicts the local MSE from the input MSE is frozen for the transfer learning.
A trainable dense layer with five nodes and elu activation are added to the architecture between the input MSE and linear node, which allows flexibility in updating the learned mask with observations.
Early stopping is still used, along with a scheduler for the learning rate, which starts off at $0.0001$ and decays to $10^{-6}$.
Examples of the masks before and after transfer learning are discussed in Section \ref{sec:wus_masks}.

\subsection{Analog Selection}\label{sec:selection}

Next, we select the best matching analogs for a given prediction target region and lead time.
We present three selection methods: global mask, regional mask, and learned mask.
All three methods select analogs by calculating the spatial-average MSE between the initialization state (observations) and all the states in the library of potential analogs.
Note, for years before $1956$, the BEST dataset does not have data at every global location.
The library of potential analogs is the CMIP6 data outlined in Table \ref{table:cmip6}, which is preprocessed using the same steps outlined in Section \ref{sec:data}, without splitting into multiple libraries.
The three methods differ in the spatial weighting of the MSE calculation.
For the global mask, the MSE is unweighted, with every grid cell on the globe contributing equally.
For the regional mask, the MSE is weighted equally over the prediction target region and zero outside the region.
For our learned mask, the MSE is weighted according to the learned mask of weights described in Section \ref{sec:mask}.

The known future of the lowest $N$ MSE potential analogs are then used as an ensemble of predictions.
For example, for a prediction with a lead time of five years, we find the $N$ lowest MSE model-states: best matching models $M$, members $m$, and years $Y$.
We then compile the prediction ensemble from five years in the future of those model-states: the same models $M$ and members $m$, but for years $Y+5$.
One of the strengths of analog methods is the freedom to choose $N$ for a given prediction task.

Instead of matching on only the initialization state, we allow for additional matching on years preceding the initialization state (which we call ``tethering'').
The motivation behind tethering is to avoid by-chance matches that have very different past and future evolutions by incorporating the time tendency in the analog selection.
The initialization state is given a weight of one, while previous years are inverse weighted such that the most recent year is the most heavily weighted.
We show the change in performance for tethers of 1, 2, and 3 years for the five regions (averaged over lead times of 1 to 10 years) in Supplementary Section S1.
In general, the regional mask has the largest improvements from increasing the tether length, while the changes in the global mask are small and inconsistent, with performance degraded in some regions and improved in others.
The learned mask generally improves, but with a smaller magnitude than the regional mask.
Throughout, we use a tether length of two years for all three analog methods (learned, global, and regional), though the optimal tether length will likely be region and lead time specific, and could be optimized further.
We leave this for future work.

\subsection{Evaluation}\label{sec:metrics}

In order to comprehensively evaluate the performance of our learned mask analogs, we use both deterministic (mean squared error, MSE) and probablistic (continuous ranked probability score, CRPS; \cite{Gneiting2007}) metrics \cite{goddard2013}.
The CRPS measures how representative a predicted distribution is, given a true value, allowing us to quantify the performance of the analog distribution.
Lower values of CRPS indicate better performance.

In Section \ref{sec:results}, we compare the performance of our learned mask analogs to global mask and regional mask analogs, in addition to the CMIP6 library (Table \ref{table:cmip6}) and four IESMs (Table \ref{table:iesms}).
For analog predictions, MSE is calculated using the mean of the top $N$ analogs, while CRPS uses the distribution of the top $N$ analogs, where $N$ is stated for each analysis.
Performance of the three analog methods versus $N$ is shown in Supplementary Section S2.
CMIP6 and IESM predictions are the mean (MSE) or distribution (CRPS) over all available members for the predicted year.

We calculate skill scores for these five prediction methods using a climatology reference.
For MSE, the climatology predictions are averages of BEST over the $30$ years immediately preceding and including the intialization year.
For CRPS, the climatology predictions are the distribution of the $30$ years of BEST data immediately preceding and including the intialization year.
For example, for a $10$ year lead time prediction initialized on $2010$, the MSE climatology is the $1981-2010$ average and the CRPS climatology is the set of samples covering $1981-2010$.
CRPS and MSE skill scores are presented as either the regional mean over time and space (lead time specific performance) or the mean over time and lead time (regional spatial performance).

\section{Results}\label{sec:results}

\begin{figure}[h!]
    \noindent\includegraphics[width=\textwidth]{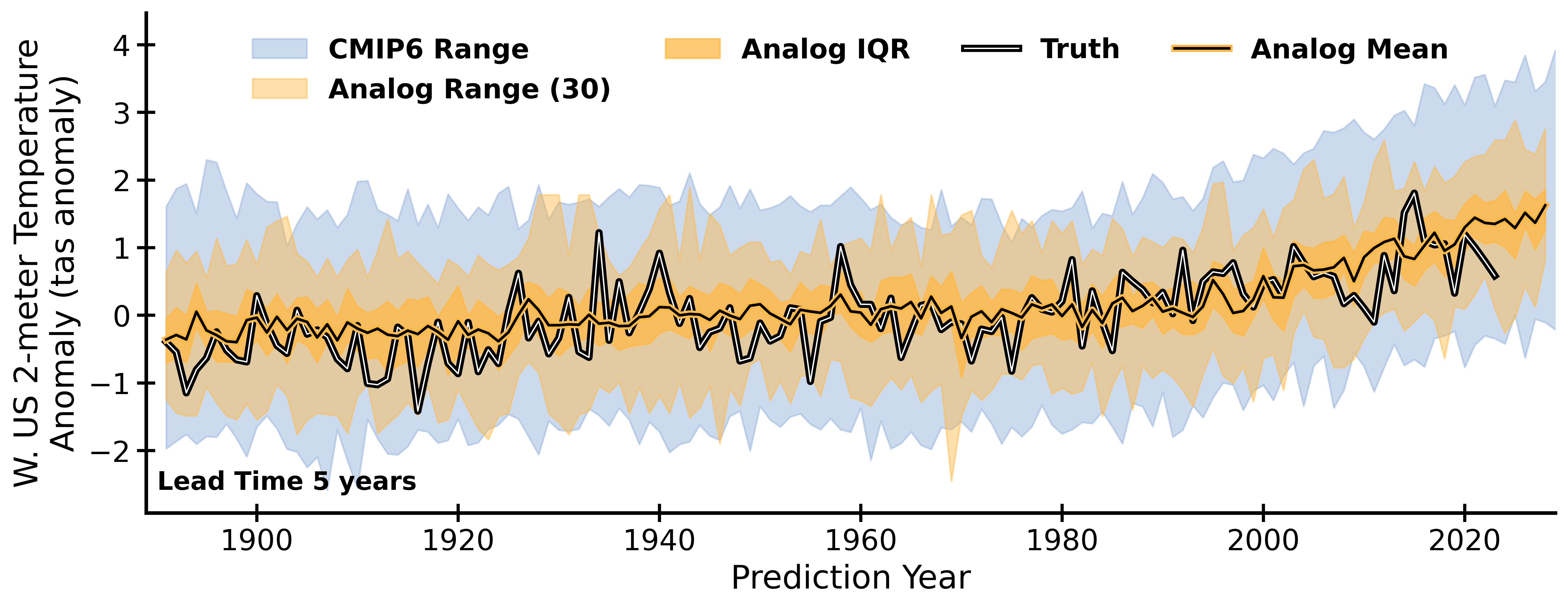}
    \caption{Example region-averaged (western US) analog predictions using the top 30 analogs for a lead time of 5 years for 2-meter temperature anomalies.
    The mean over the 30 analogs (black-yellow line), interquartile range (dark yellow shade) and full range (light yellow shade) are shown for comparison with the CMIP6 range (blue shade) and observations (BEST, black-white line).
    }
    \label{fig:wus_timeseries}
\end{figure}

Each prediction presented here is for a single annual-average of the target region map, e.g., the 5 year prediction initialized in 2010 is for the year 2015.
For each region and lead time, analog method predictions are made for each year in the stated time range and combined into a time series.
An example prediction made by our learned mask analog method for 2-meter temperature for the western United States with a $5$ year lead time is shown in Figure \ref{fig:wus_timeseries}.
The full range, interquartile range (IQR), and mean of the top 30 analogs are shown, along with the BEST data (truth) and full range of CMIP6 simulations.
We quantify the skill of our approach compared to these others throughout this section, but note that a key strength of analog predictions is the narrowing of the range relative to the full set of CMIP6. 

\subsection{Learned Masks}\label{sec:wus_masks}

\begin{figure}[h!]
    \centering
    \noindent\includegraphics[width=0.49\textwidth]{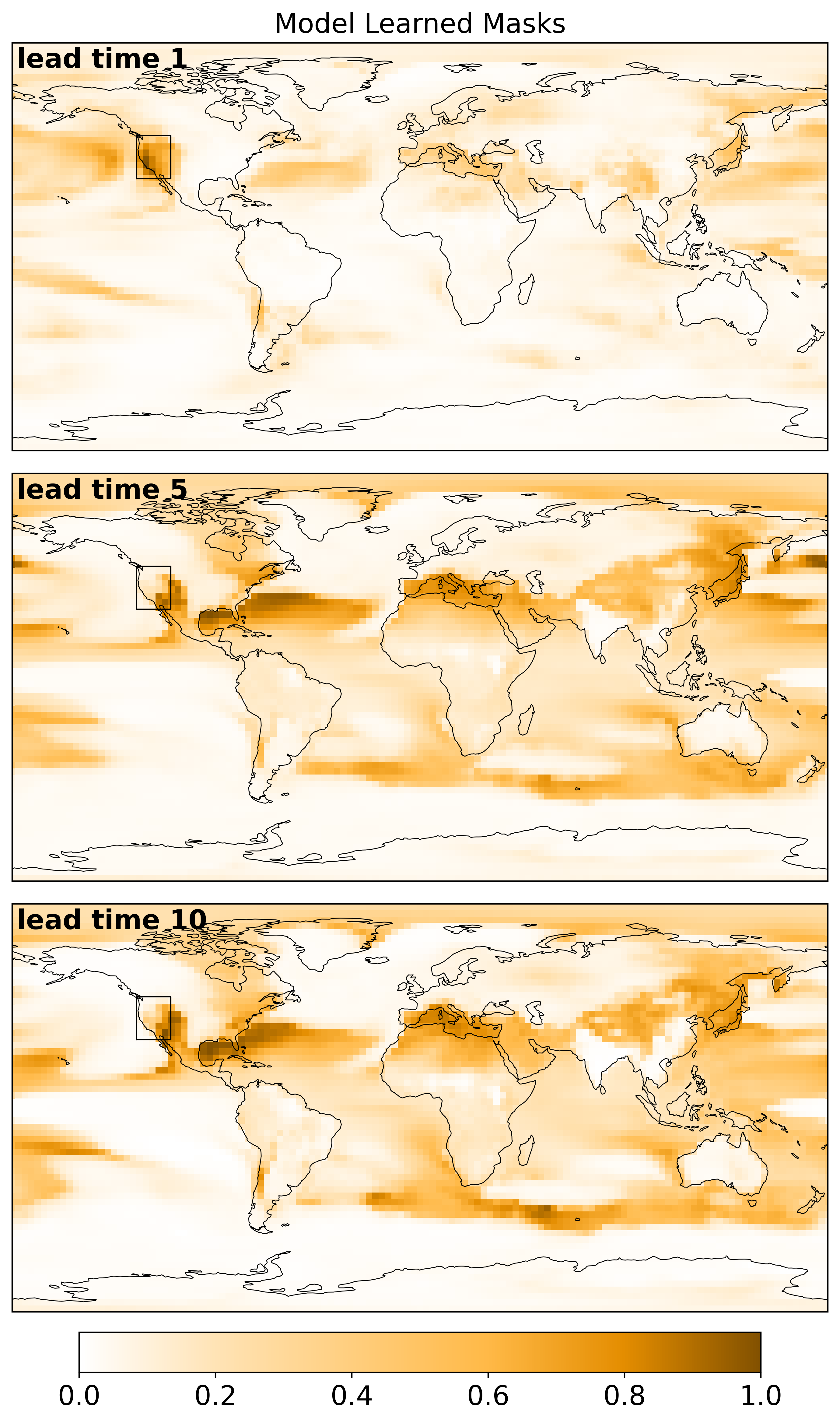}
    \noindent\includegraphics[width=0.49\textwidth]{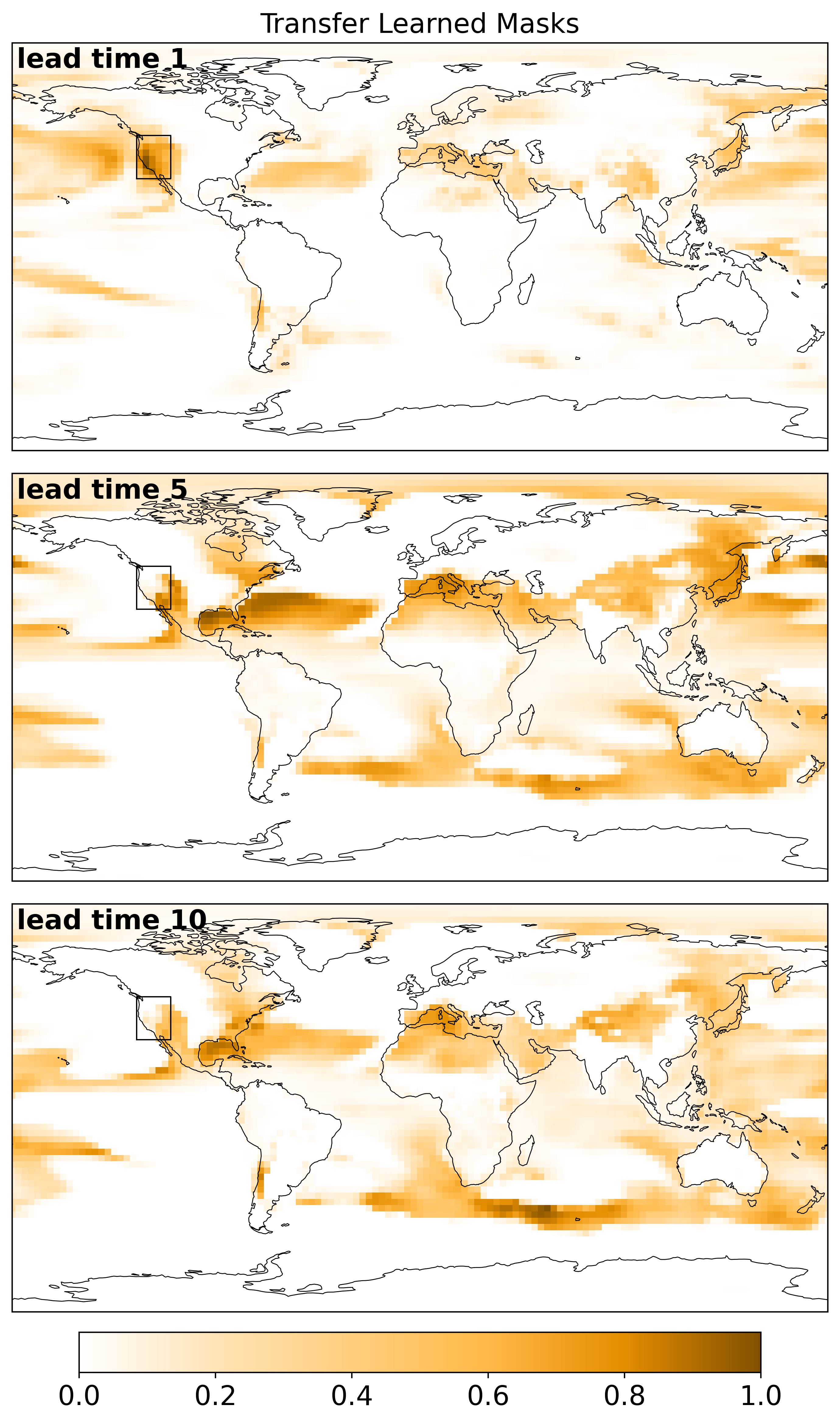}
    \caption{Learned masks for the western United States region (black bounded area), scaled by the maximum weight to highlight the pattern.
    The left panels show model learned masks for lead times of 1, 5, and 10 years.
    The right panels show transfer learned masks (see Section \ref{sec:mask}) for the same lead times.
    }
    \label{fig:wus_masks}
\end{figure}

We begin by highlighting the interpretability of the learned mask method.
The learned mask for the predictions shown in Figure \ref{fig:wus_timeseries} is shown in the middle row of the left column of Figure \ref{fig:wus_masks}, along with the learned masks for lead times of 1 and 10 years.
The right column shows the refined masks after finetuning via transfer learning on a limited subset of the BEST dataset.
The learned masks for all the lead times are shown in Figures S5 and S6 in Supplemental Section S3.
The change in prediction skill is negligible when using the transfer learned mask instead of the model learned masks, as can be seen in Supplementary Section S4.
Thus, while transfer learning appears to create a sparser learned mask, it does not appear to improve skill on the limited observational testing set.
This may indicate that there are not enough samples in the BEST training set to overturn any learned model biases, or that the model biases are negligible.
Although some studies suggest finetuning on observations can improve results \cite{Martin2025}, other works suggest we may not have enough data to see improvement \cite{Mayer2025}.
For this reason, we drop the finetuning step and focus on the learned masks trained on the CMIP6 library alone (left column of Figure \ref{fig:wus_masks}).

We now discuss some of the patterns and consistent features in the learned masks.
Longer lead times (Figure \ref{fig:wus_masks}; Supplemental Figure S5) have more important features, i.e. the masks become less sparse with lead time.
In particular, more distant features from the target region increase in importance for longer lead times as the relative importance of the target region decreases.
This behavior is true for the learned masks across the other four target regions as well (Supplemental Section S3). 

A commonality across the learned masks for all target regions is that the oceans almost always have more weight than the land, reflecting the longer memory (and higher heat capacity) of the oceans and its associated importance in multi-year prediction \cite{Meehl2014, Meehl2021_memory}.
As an example, the Atlantic off the southeastern US and extending eastward often has large importance.
Coupled with the importance of the Gulf of Mexico, this could indicate the importance of the Gulf Stream for skillful multi-year predictions.
This signal is especially clear for the northern Europe target region (Supplemental Figure S9). 

We find evidence of a shift in the mask weights from focusing on features important for internal variability, to features important for longer term warming trends.
For example, the El Nino Southern Oscillation (ENSO) is an important driver of $1-2$ year predictability of surface temperature across the globe \cite{Meehl2021_memory}.
The warm tongue over the eastern tropical Pacific shows some importance for the $1$ year lead time for the Amazon, southern India, and the African Great Lakes, but is absent by year $2$ for these regions and for all lead times for the western U.S. and northern Europe.

At longer lead times, several areas consistently appear for all target regions and may be markers of a robust warming trend \cite{diffenbaugh2023, barnes2024}.
These include the Tibetan Plateau, the western US, the Mediterranean, north Africa, and oceanic features off the southwest coast of Africa and southeast coast of Australia.
Many of these features are locations with documented larger climate change signal-to-noise ratios across CMIP6 models that act as robust identifiers of climate change \cite{barnes2020}.

\subsection{Analog Methods Comparison}\label{sec:analogs}

We start our method comparison with an analysis of the CMIP6 library and the three analog methods: the learned mask, global mask, and regional mask.
All results in this section are averaged over the period $1931-2023$, which is the time period for which all BEST grid cells evaluated here are available.
All analog methods use the top 30 analogs.

\begin{figure}[h!]
    \centering
    \noindent\includegraphics[width=0.9\textwidth]{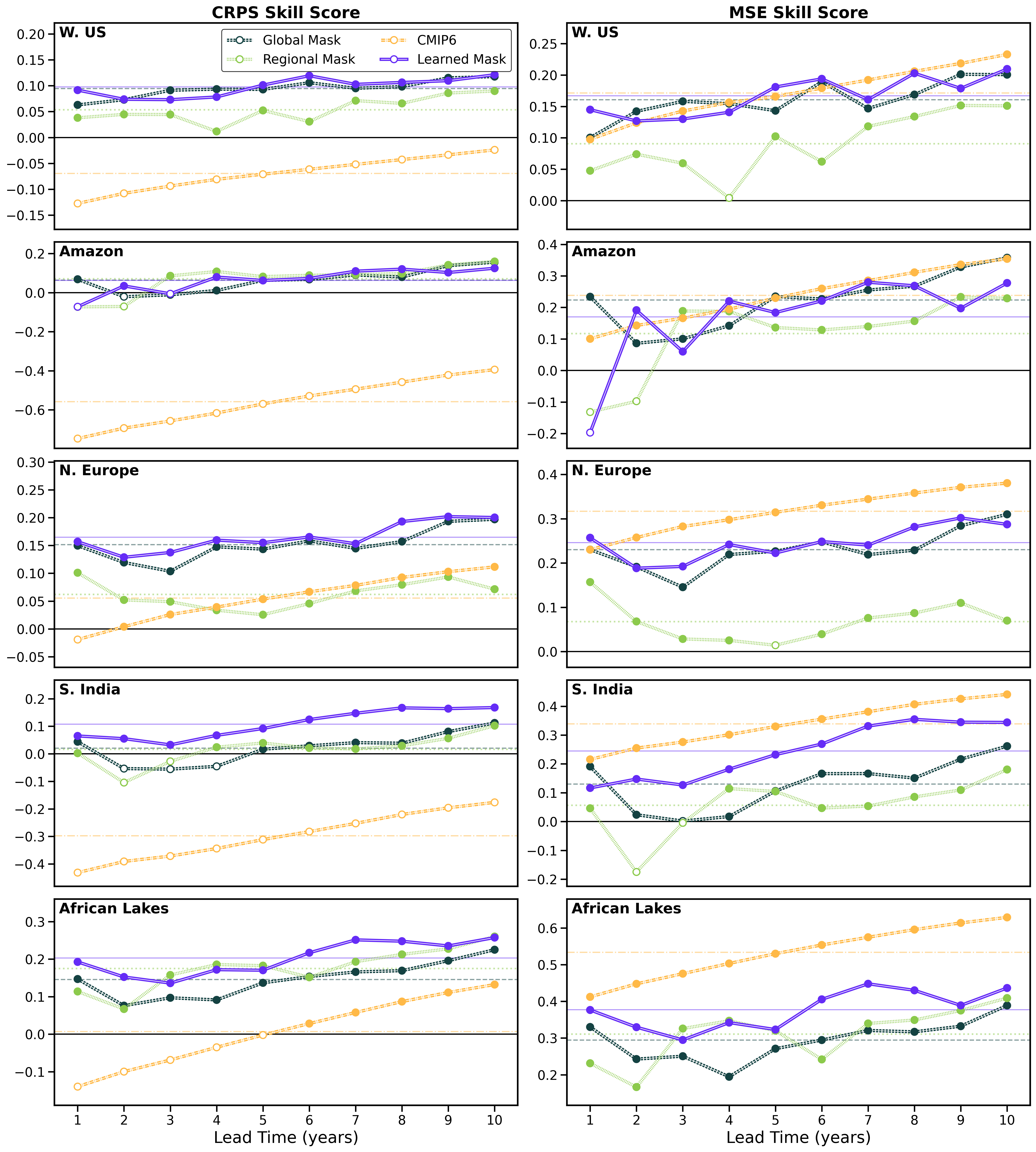}
    \caption{CRPS (left) and MSE (right) skill scores as a function of lead time.
    The mean over lead times for each method is shown as a horizontal line, as is the zero-line.
    Analog methods use the top 30 analogs, while CMIP6 uses the full set of members (see Table \ref{table:cmip6}).
    Skill is relative to a 30-year climatology and the metrics are averaged over the time period 1931-2023 (93 years).
    }
    \label{fig:full_lead_skill}
\end{figure}

Figure \ref{fig:full_lead_skill} shows the CRPS (left) and MSE (right) skill scores for lead times 1-10 years for the five target regions.
Filled circles for all methods in Figure \ref{fig:full_lead_skill} indicate that the metric is lower than both the reference climatology (positive skill) and the 5th percentile of the bootstrapped climatology.
The bootstrapping method is described in Supplemental Section S5.

There is a general trend of increasing skill with lead time in Figure \ref{fig:full_lead_skill}.
This is due to the reference climatology degrading in performance with lead time more sharply than the other prediction methods.
This is most obvious for the CMIP6 predictions, which do not depend on the lead time (the CMIP6 prediction for $2020$ is the ensemble of model outputs for $2020$, regardless of the initialization year).
This can be seen in Supplementary Section S5, which shows the raw metrics for the reference climatology alongside the learned mask analog predictions and CMIP6 library.

The most obvious result in Figure \ref{fig:full_lead_skill} is the lack of CRPS skill for the CMIP6 library, which has negative skill for all lead times for the western U.S., Amazon, and southern India, and the lowest average CRPS skill score (denoted by the horizontal lines in each panel) for all regions.
Conversely, the CMIP6 library produces the highest average MSE skill score for all five regions, and is especially skillful at longer lead times.
Low CRPS skill can be the result of a mean bias (the distribution is shifted away from the true mean), from overdispersion (the distribution is too wide), or from a combination of a mean bias and underdispersion or overdispersion.
The CMIP6 results indicate that the CMIP6 ensemble mean captures the trend well, but the CMIP6 distribution is overdispersed.

The regional mask method is the worst or near-worst in terms of average MSE skill score for all regions, though it does show positive skill for most cases.
The regional mask does better in terms of CRPS skill, but is worse than the learned mask analog in all regions except the Amazon.
These results indicate that the regional mask ensemble mean is biased (lower MSE skill), which drives the CRPS skill lower than the learned mask analogs.
The mean bias highlights the importance of matching outside the target region for identifying trends.
This point is reinforced by the global mask ensemble mean MSE skill, which is higher than the regional mask in four out of five regions.
This translates to a higher CRPS skill compared to the regional mask in some regions, but in others (S. India and the Amazon) the global mask CRPS skill is similar or lower than the regional mask.
This indicates the importance of the target region, which is downweighted when matching everywhere on the globe equally, for generating ensembles with appropriate ensemble dispersion.

The learned mask predictions show overall better ensemble mean MSE skill than the other analog methods in four out of five regions, and overall similar or better CRPS skill in all regions.
This indicates that learned mask analogs, by highlighting important matching areas for the specific prediction task, have the strengths of both global and regional analog methods.

We report two additional summary statistics: the mean skill and the positive skill fraction.
The mean skill is the skill score averaged over all grid cells in the five regions.
The positive skill fraction is the fraction of grid cells that have positive skill.
The mean skill over all regions and lead times for the learned mask (CRPS: 0.128, MSE: 0.256) is higher than both the global mask (CRPS: 0.107, MSE: 0.251) and regional mask (CRPS: 0.087, MSE: 0.152), with the CMIP6 library last in CRPS and leading in MSE (CRPS: -0.221, MSE: 0.337).
The fraction of grid cells over all regions with positive mean skill for the learned mask (CRPS: 0.89, MSE: 0.93) is similar to the global mask (CRPS: 0.89, MSE: 0.96) and the regional mask (CRPS: 0.93, MSE: 0.96), with CMIP6 similar for MSE, but worse for CRPS (CRPS: 0.34, MSE: 0.94).
These metrics, along with a spatial map of CRPS and MSE skill scores, can be seen in Supplementary Section S6.

\subsection{IESM Comparison}\label{sec:iesms}

We now compare our learned mask analog method to a set of initialized Earth system models (IESMs, described in Table \ref{table:iesms}), as well as to the CMIP6 library.
All results in this section are averaged over the period $1994-2018$.
The shorter time period is due to the bias-correction of the IESMs and the requirement that all members have predictions for the time period.
We use the top 70 analogs (to match the 70 members from the IESMs).

\begin{figure}[h!]
    \centering
    \noindent\includegraphics[width=0.9\textwidth]{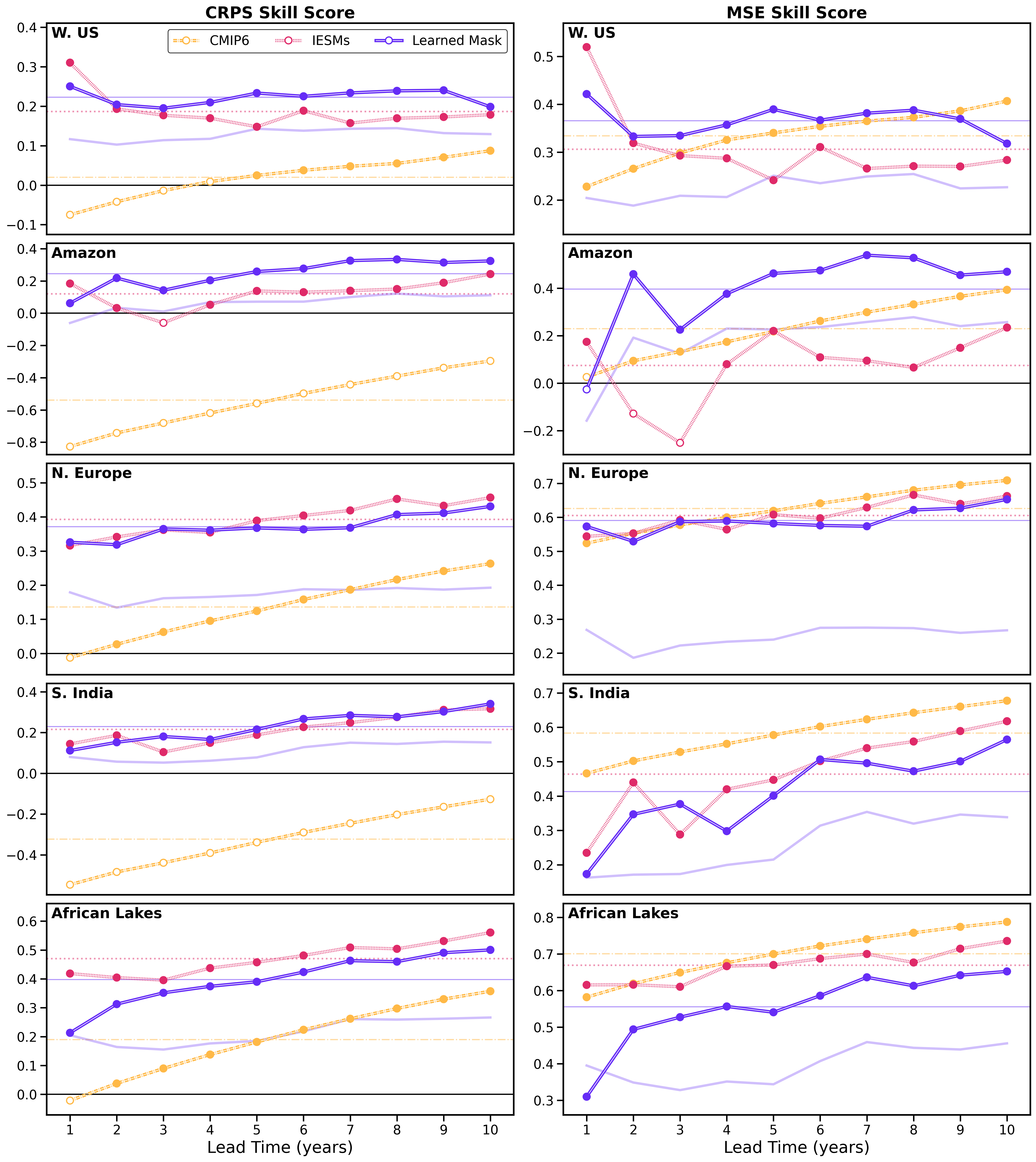}
    \caption{Same as Figure \ref{fig:full_lead_skill}, but comparing the learned mask analogs (70 analogs) to IESMs (70 members) and CMIP6 (all members) for the time period 1994-2018 (25 years).
    }
    \label{fig:iesm_lead_skill}
\end{figure}

Figure \ref{fig:iesm_lead_skill} shows the CRPS (left) and MSE (right) skill scores for lead times 1-10 years for the five target regions.
Again, the filled circles for all methods indicate that the metric is lower than both the reference climatology and the 5th percentile of the bootstrapped climatology metrics (Supplemental Section S5).

The skill scores for the learned mask analog predictions for the time period $1931-2018$ (semi-transparent lines) are also shown in Figure \ref{fig:iesm_lead_skill}.
The skill is almost always higher for both metrics in the more recent period ($1994-2018$) than the longer time period ($1931-2018$).
The increased skill in the more recent period is largely driven by the reference climatology failing to capture the trend using the most recent thirty years, i.e., the recent past becomes a weaker predictor of the future.

As with the longer time period ($1931-2023$), the CMIP6 library ensemble mean is skillful (MSE), but the distribution is too wide (lower CRPS skill).
Overall, the learned mask analogs and the IESMs perform similarly for both metrics.
Specifically, the learned mask analogs are more skillful for the western US and the Amazon beyond one year, while the IESMs are more skillful for the African Great Lakes.
Skill for southern India and northern Europe is more mixed, with the IESMs slightly better overall.

\begin{figure}[h!]
    \centering
    \noindent\includegraphics[width=0.45\textwidth]{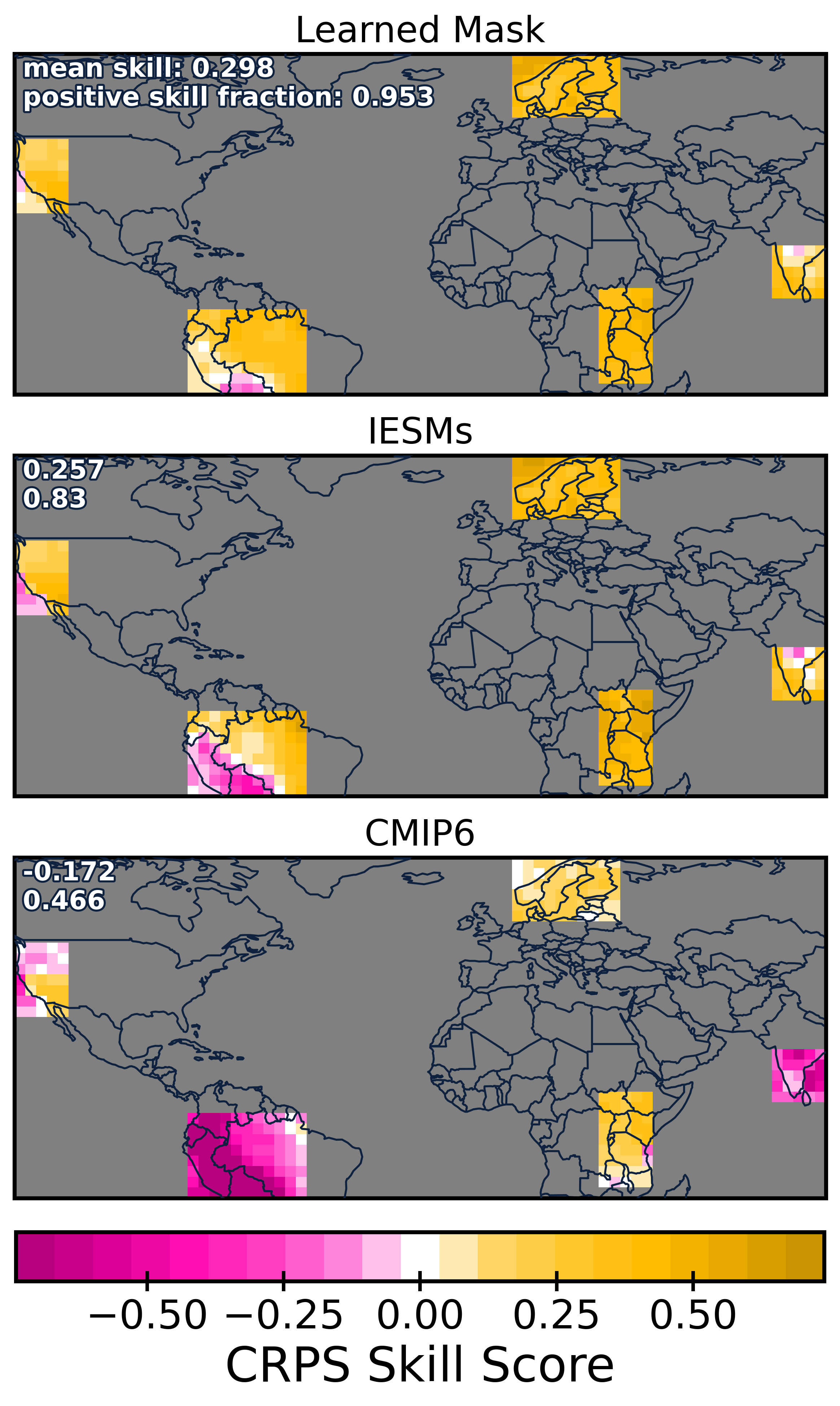}
    \noindent\includegraphics[width=0.45\textwidth]{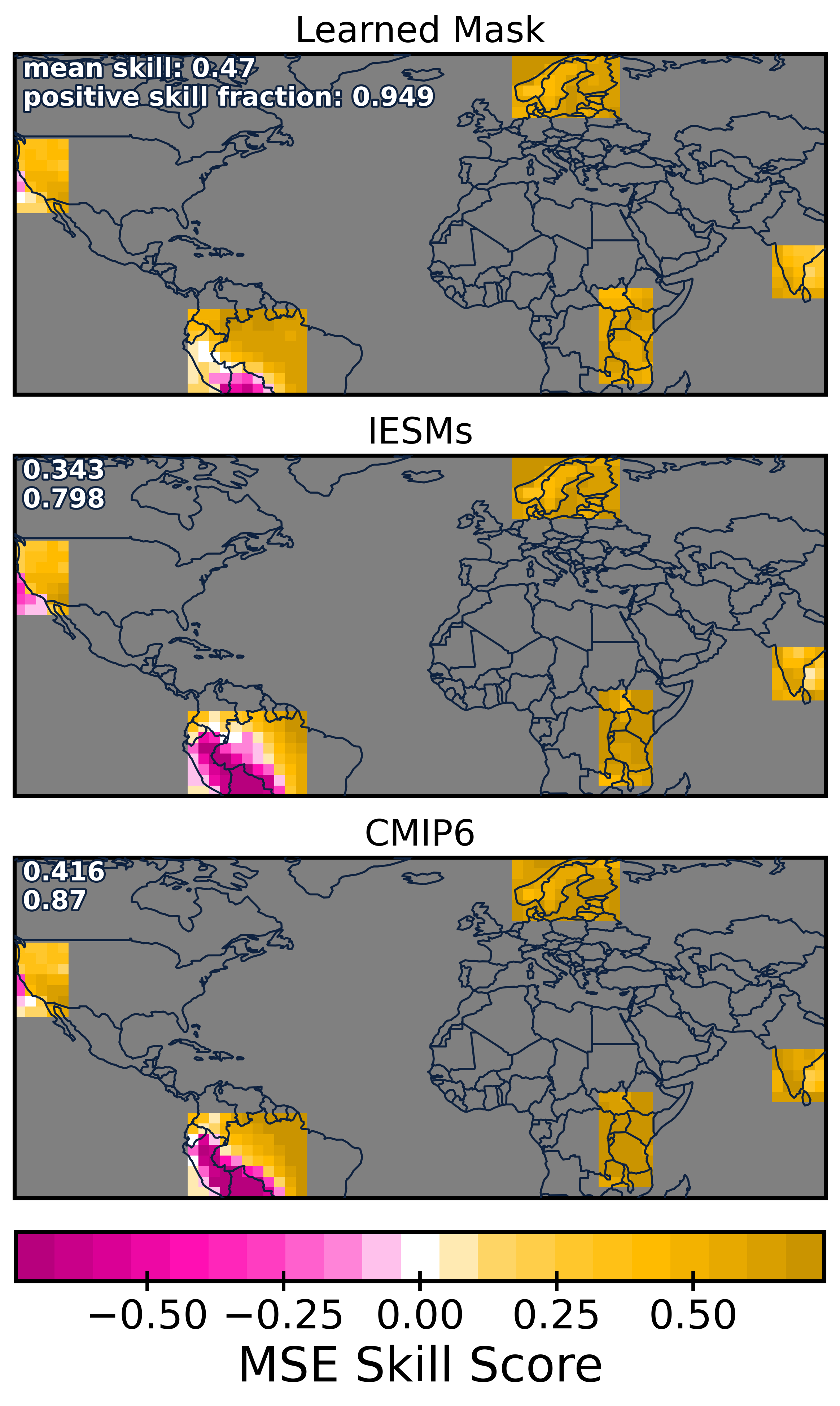}
    \caption{Spatial CRPS (left) and MSE (right) skill scores.
    Skill is averaged over lead times 1-10 years and the time period 1994-2018 (25 years).
    The mean skill over the five regions and the fraction of grid cells with positive skill are reported for each method in the upper left of each panel.
    }
    \label{fig:iesm_region_skill}
\end{figure}

Figure \ref{fig:iesm_region_skill} shows the spatial CRPS (left) and MSE (right) skill scores for the learned mask analog, IESMs, and CMIP6.
In the top left corner of each panel the mean skill and positive skill fraction are reported.
The learned mask analog predictions have higher mean MSE and CRPS skill than both the IESMs and CMIP6 library.
The learned mask analog predictions also have the highest positive skill fraction.
In particular, the learned mask predictions for the Amazon and western US regions show improvement in the spatial coverage of positive MSE and CRPS skill, with MSE skill improvement for southern India as well.

\section{Conclusions}\label{sec:conclusions}

We have presented a model-analog framework for multi-year-to-decadal prediction of regional 2-meter temperature.
The framework incorporates a machine learned mask of weights that highlights important precursors for use in model-analog forecasts.
Using deterministic and probabilistic metrics, we have shown across five regions and lead times of one to ten years that our learned mask analog method:
\begin{itemize}
    \item has higher average MSE (ensemble mean) and CRPS (distribution) skill than global and regional mask analog methods,
    \item has higher CRPS skill than the CMIP6 library for all regions and lead times, which is a result of improved prediction of distribution spread,
    \item has similar overall skill to a set of four initialized Earth system models (70 members total) in both CRPS and MSE.
\end{itemize}

Our machine learning model-analog framework can be further improved.
In this work, we explored matching on and predicting the same variable, though multiple matching variables is likely to improve skill.
We used tethering (matching on multiple years) and explored a learned mask that was refined using transfer learning, but did not optimize either of these processes.
Expanding the library of potential analogs can lead to improved matching and skill: we used a multi-model library of CMIP6 simulations as potential analogs, but analog methods allow for incorporating any relevant existing dataset into the potential analog library.

There are several benefits inherent to model-analog prediction methods.
Analog predictions do not require expensive model runs, beyond what is currently available.
Analog methods allow trivial adjustments to ensemble size, thus they can produce flexible ensembles that can be used for probabilistic prediction of extremes.
Analog predictions avoid the issue of initialization shock and climate drift faced by initialized Earth system models, potentially reducing the need for bias corrections \cite{meehl2021, meehl2022}, though analogs still assume that the models used in the analog library evolve consistently with observations.

In addition to overall higher skill than other analog methods, the learned mask of precursor weights makes our prediction framework interpretable.
We analyzed our learned masks and found that the areas of importance often aligned with known climate drivers, such as the El Nino Southern Oscillation, the Gulf Stream, and known indicators of a warming trend.
Analyses such as this are a strength of our interpretable approach, allowing us to identify predictable signals/patterns and connect them to physical understanding.

\section*{Data availability statement}
Berkeley Earth Surface Temperature \cite{best} can be obtained from \url{https://berkeleyearth.org/data/}, CMIP6 data and initialized Earth system model runs can be obtained from the Earth System Federation Grid \cite{esgf} at \url{https://esgf.github.io/index.html}.
Code underlying the machine learning model-analog framework presented is available at \url{https://github.com/mafern/mask-analog-predictions}.
\ack
Funding for this project was provided, in part, by the Regional and Global Model Analysis program area of the U.S. Department of Energy's (DOE) Office of Biological and Environmental Research (BER) as part of the Program for Coordinated Model Diagnosis and Intercomparison (PCMDI) and by the Global Water Security Center through DoD HQ00342420014.
\section*{References}
\bibliographystyle{unsrt}
\bibliography{references}
\renewcommand\thesection{S\arabic{section}}
\renewcommand\thefigure{S\arabic{figure}}
\setcounter{figure}{0}
\setcounter{section}{0}

\doublespacing
\begin{center}
    {\LARGE \bf Supplementary Material}
\end{center}
\onehalfspacing



\section{Tether Length Comparison}\label{sec:tether_comp}

\begin{figure}[!ht]
    \centering
    \includegraphics[width=0.78\textwidth]{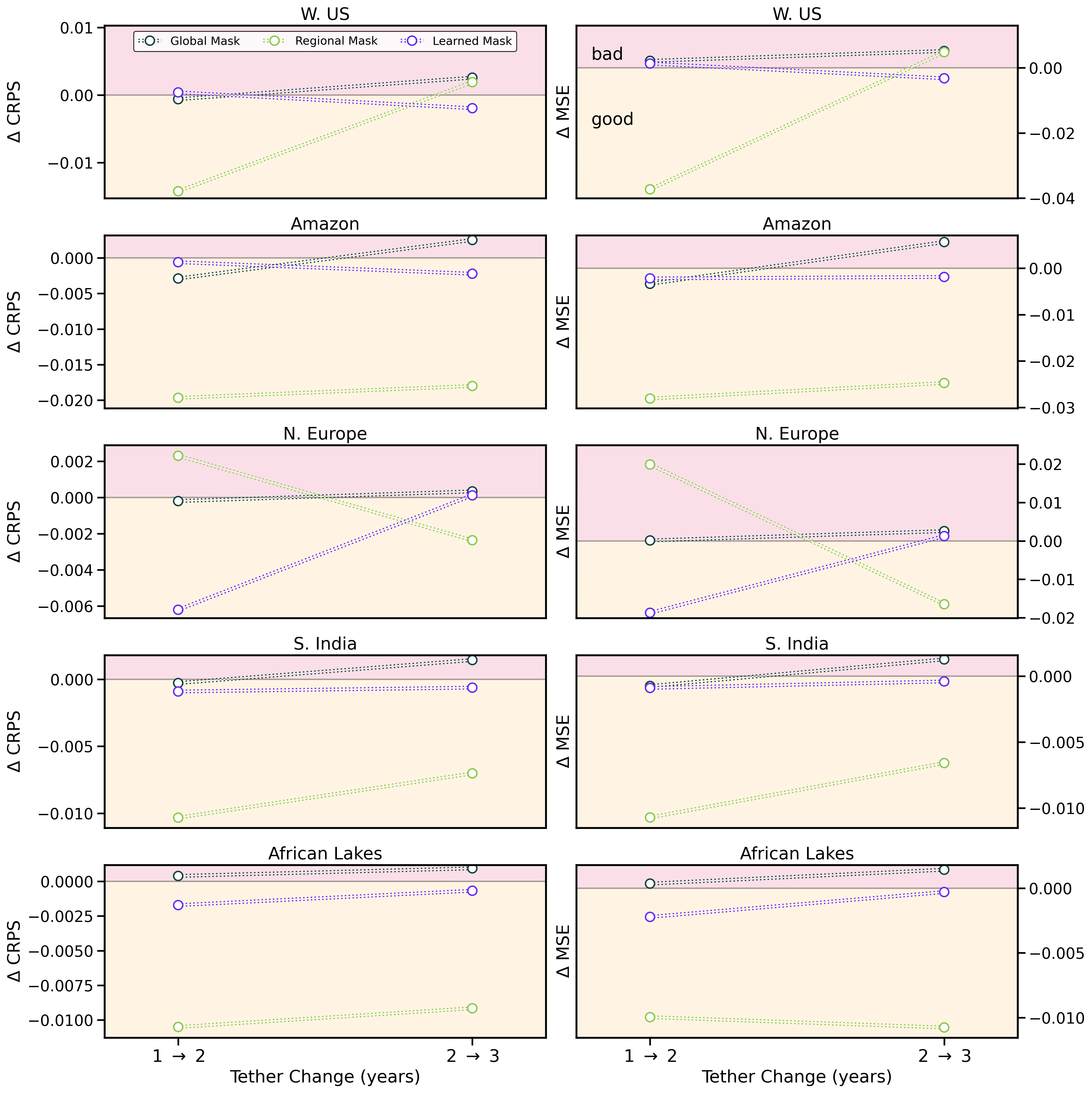}
    \caption[Metric comparison betwee tether lengths, 1931-2023]{
        Change in CRPS (left) and MSE (right) for the five regions defined in the manuscript, for 1931-2023.
        Within each panel, the left point ($1\to2$) is the change in the metric from predictions made with a tether length of 1 year to a tether length of 2 years.
        Thus, a negative value means that the longer tether length improves the prediction (reduces the CRPS or MSE).
        Likewise, the right point ($2\to3$) shows the change from a 2-year tether to a 3-year tether.
    }
    \label{fig:tether_comp_full}
\end{figure}

\begin{figure}[!ht]
    \centering
    \includegraphics[width=0.78\textwidth]{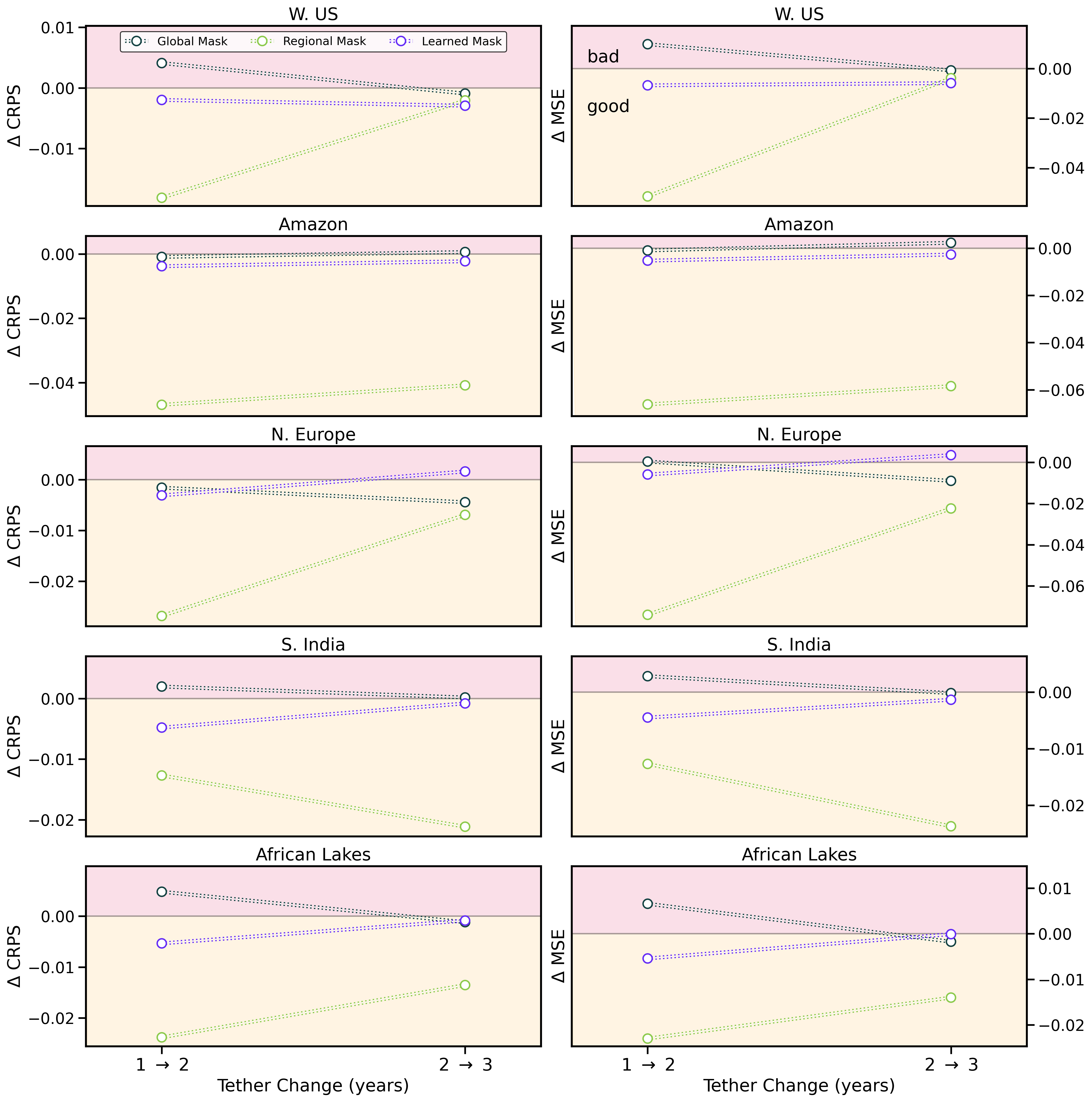}
    \caption[Metric comparison betwee tether lengths, 1994-2018]{
        Same as Figure \ref{fig:tether_comp_full}, but for the time period 1994-2018.
    }
    \label{fig:tether_comp_iesm}
\end{figure}
\clearpage


\section{Metrics Versus Number of Analogs}\label{sec:number_analogs}

\begin{figure}[!ht]
    \centering
    \noindent\includegraphics[width=0.8\textwidth]{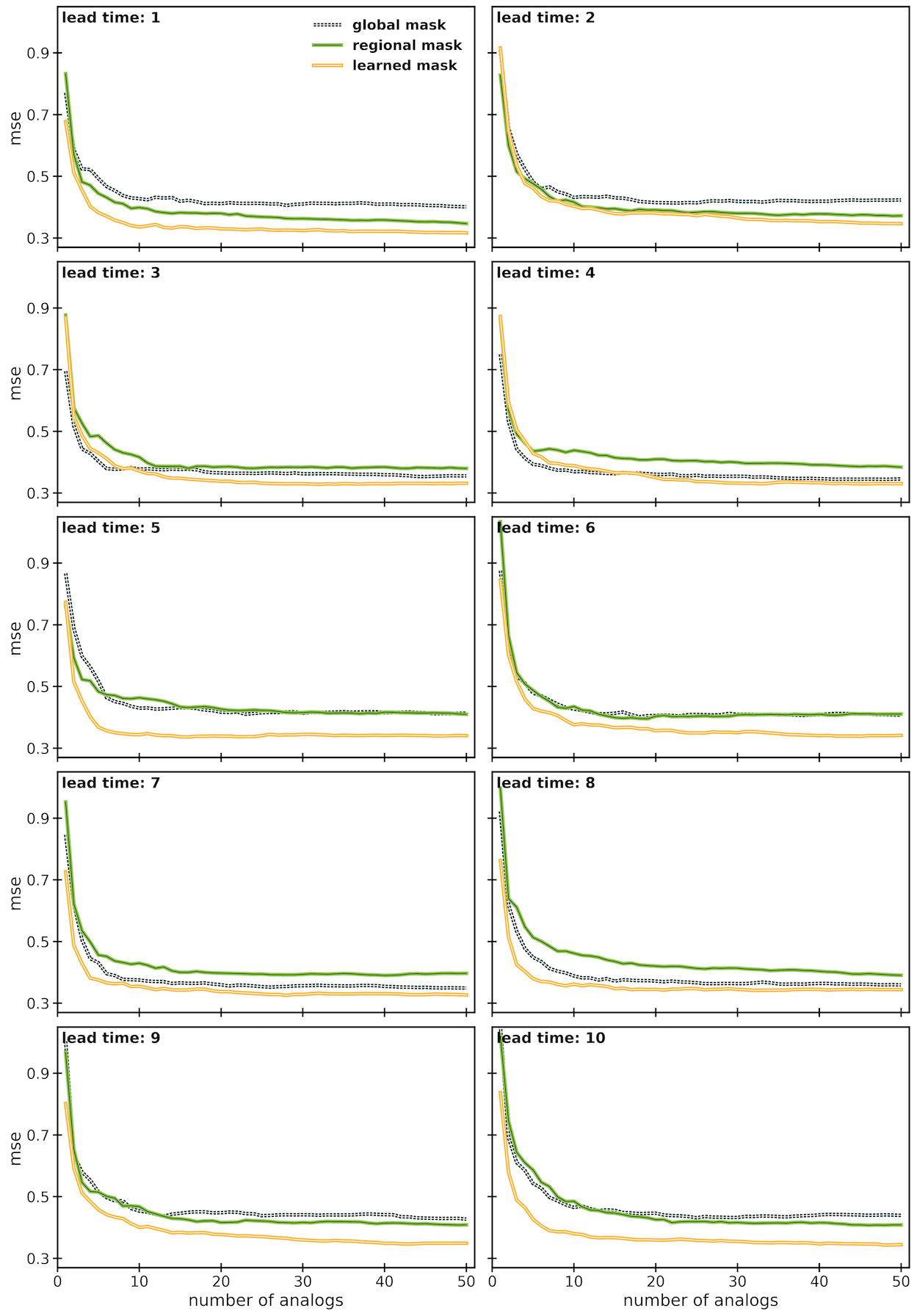}
    \caption[MSE versus number of analogs]{
        MSE versus number analogs used to calculate the mean prediction, for all lead times, for the western United States.
        Also shown are the global and regional mask results.
    }
    \label{fig:mse_nana_full}
\end{figure}

\begin{figure}[!ht]
    \centering
    \noindent\includegraphics[width=0.8\textwidth]{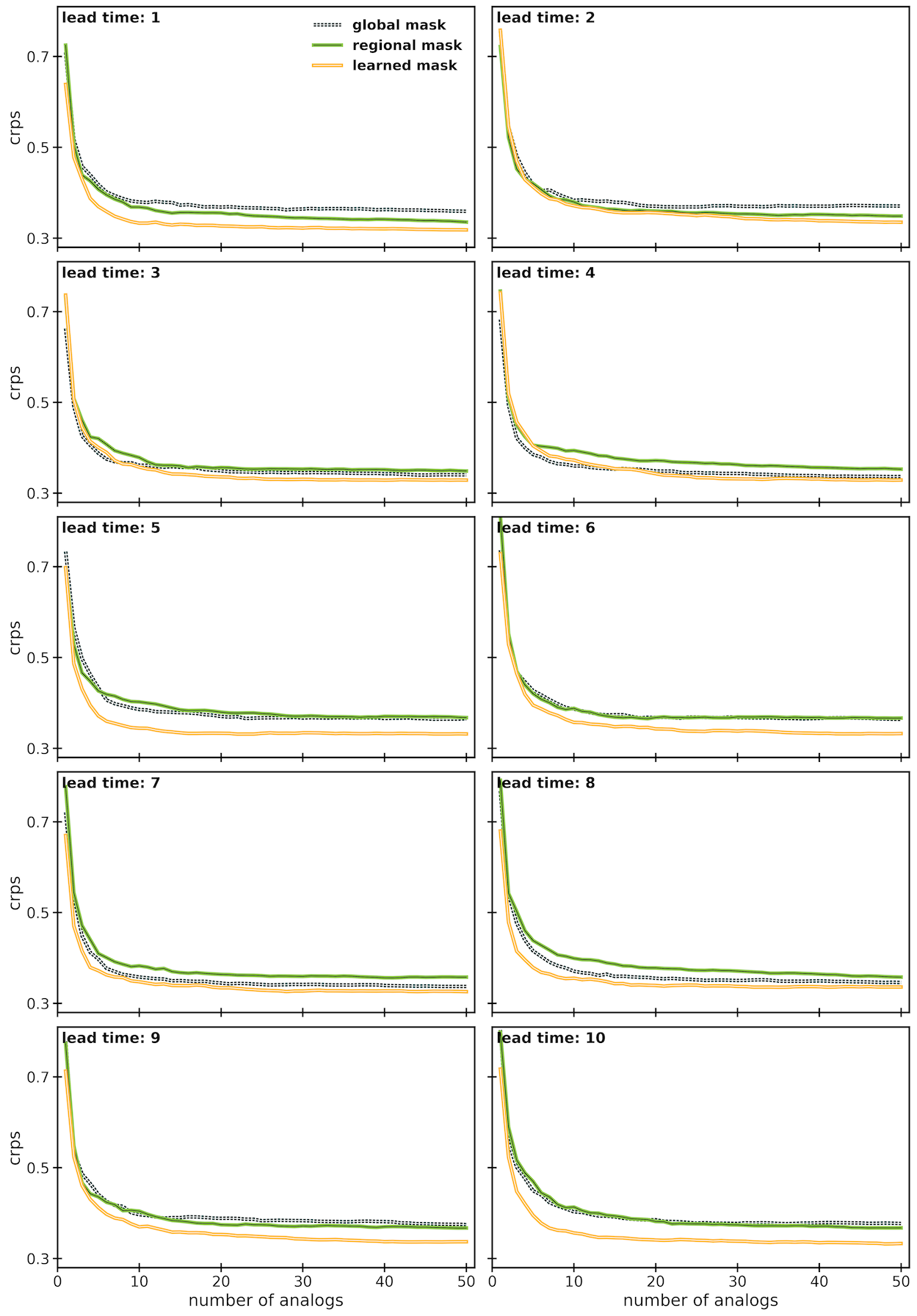}
    \caption[CRPS versus number of analogs]{
        Same as Figure \ref{fig:mse_nana_full}, but for CRPS.
    }
    \label{fig:crps_nana_full}
\end{figure}
\clearpage


\section{All Learned Masks}\label{sec:masks}

\begin{figure}[!ht]
    \centering
    \noindent\includegraphics[width=0.85\textwidth]{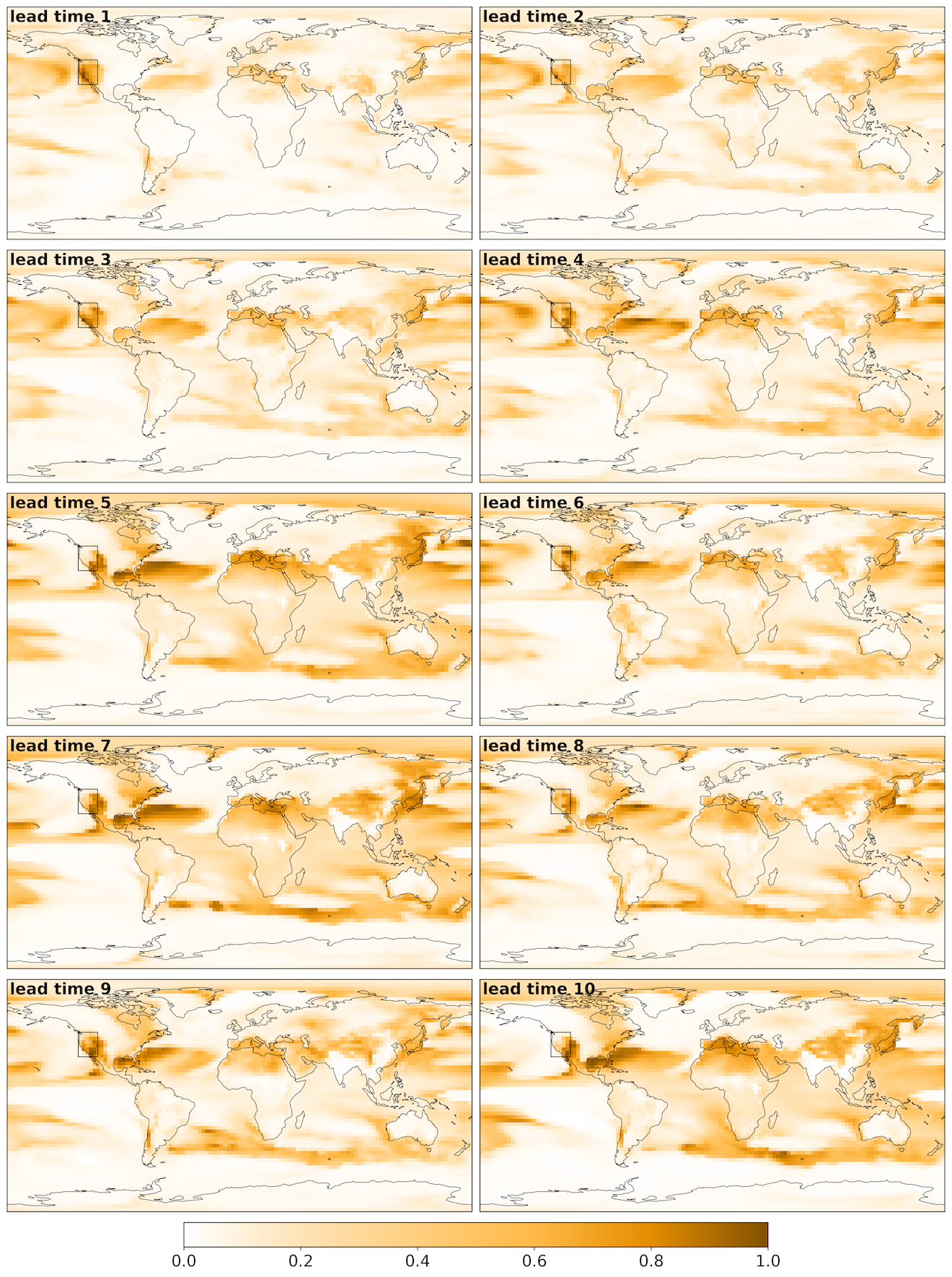}
    \caption[Western United States learned masks, 1-10 years]{
        Model learned masks for the western United States region for lead times 1-10 years.
    }
    \label{fig:wus_all_learned_masks}
\end{figure}

\begin{figure}[!ht]
    \centering
    \noindent\includegraphics[width=0.85\textwidth]{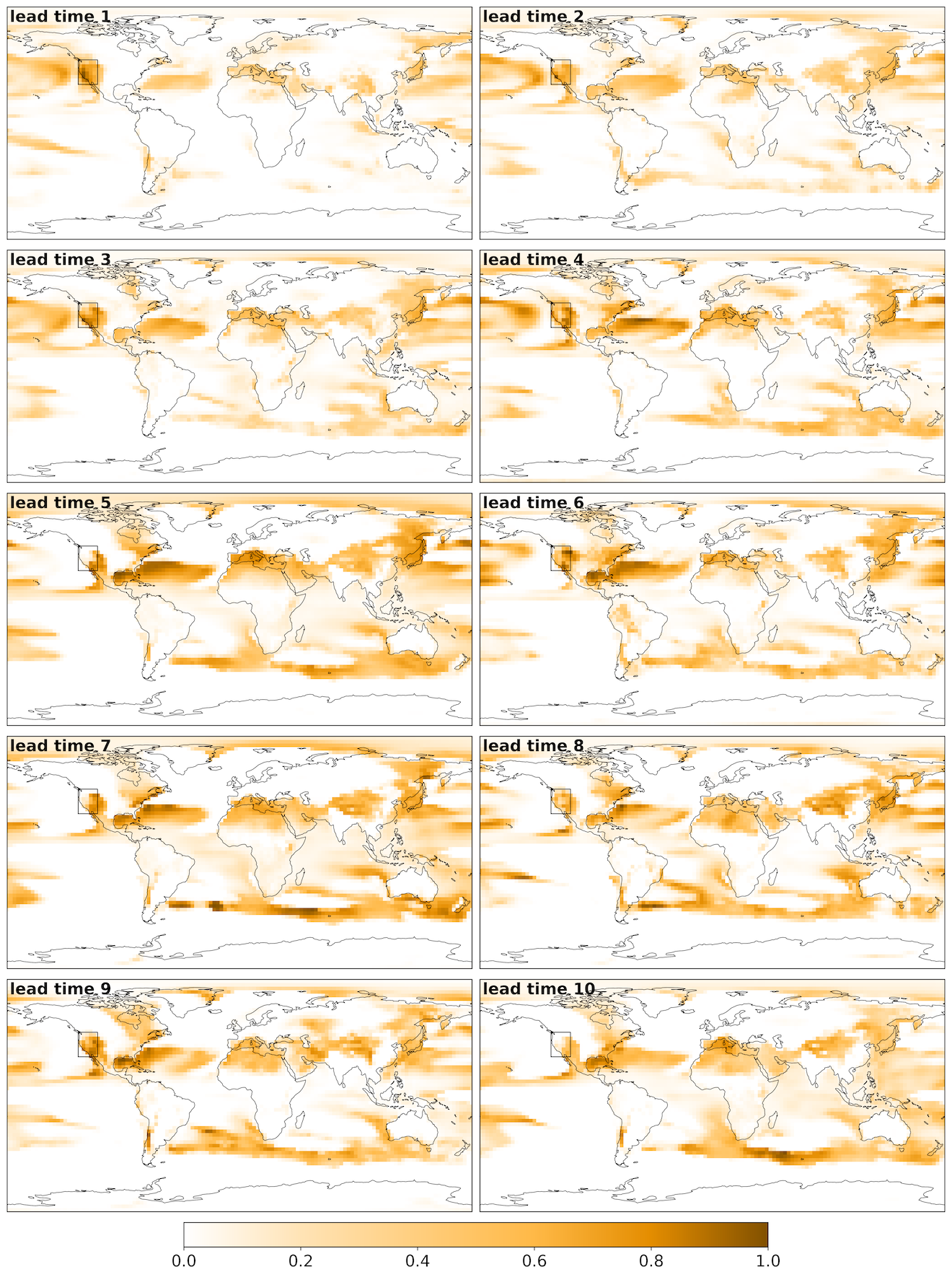}
    \caption[Western United States transfer learned masks, 1-10 years]{
        Transfer learned masks for the western United States region for lead times 1-10 years.
    }
    \label{fig:wus_all_tl_masks}
\end{figure}

\begin{figure}[!ht]
    \centering
    \noindent\includegraphics[width=0.85\textwidth]{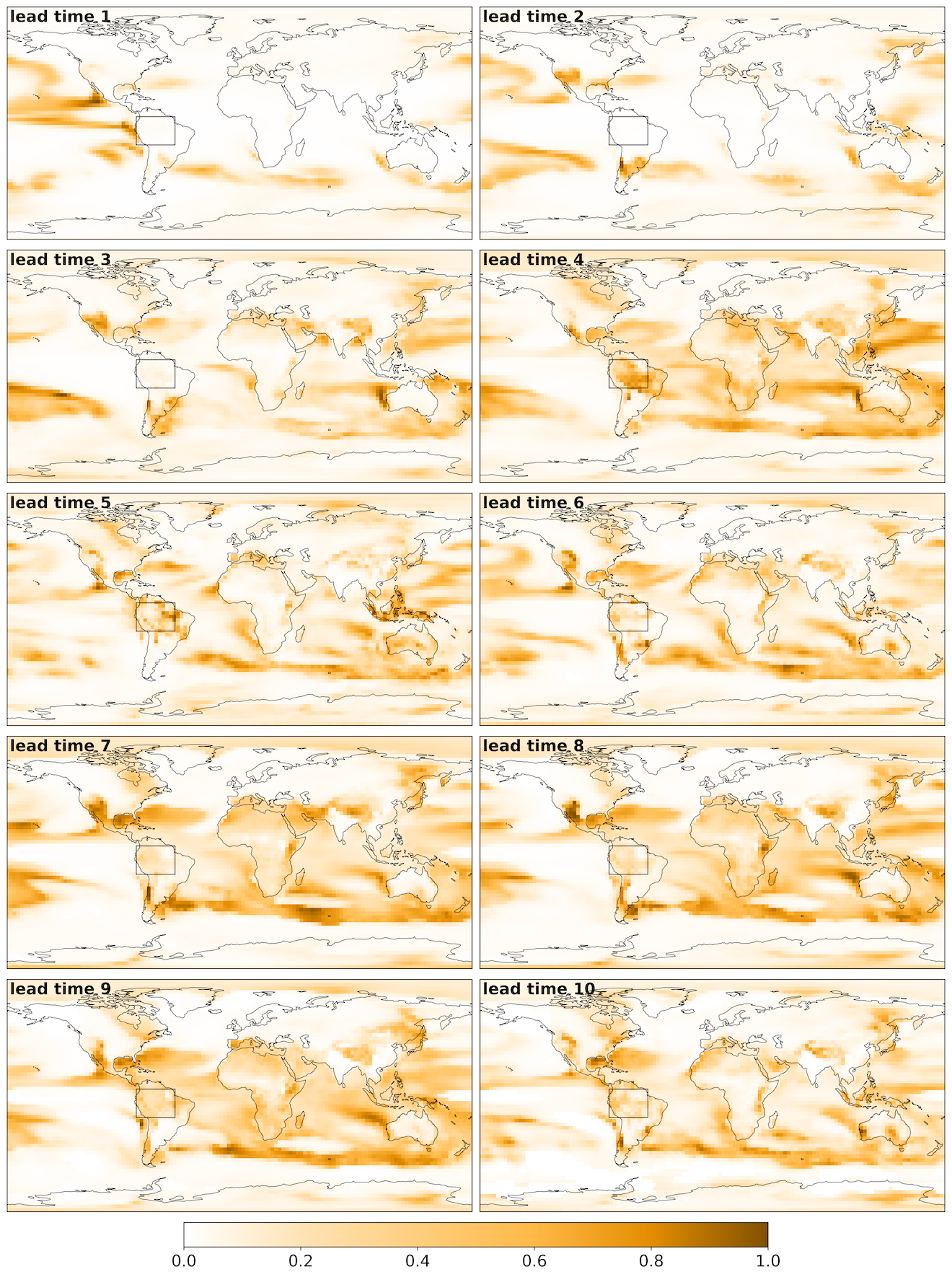}
    \caption[Amazon learned masks, 1-10 years]{
        Model learned masks for the Amazon region for lead times 1-10 years.
    }
    \label{fig:amzn_all_learned_masks}
\end{figure}

\begin{figure}[!ht]
    \centering
    \noindent\includegraphics[width=0.85\textwidth]{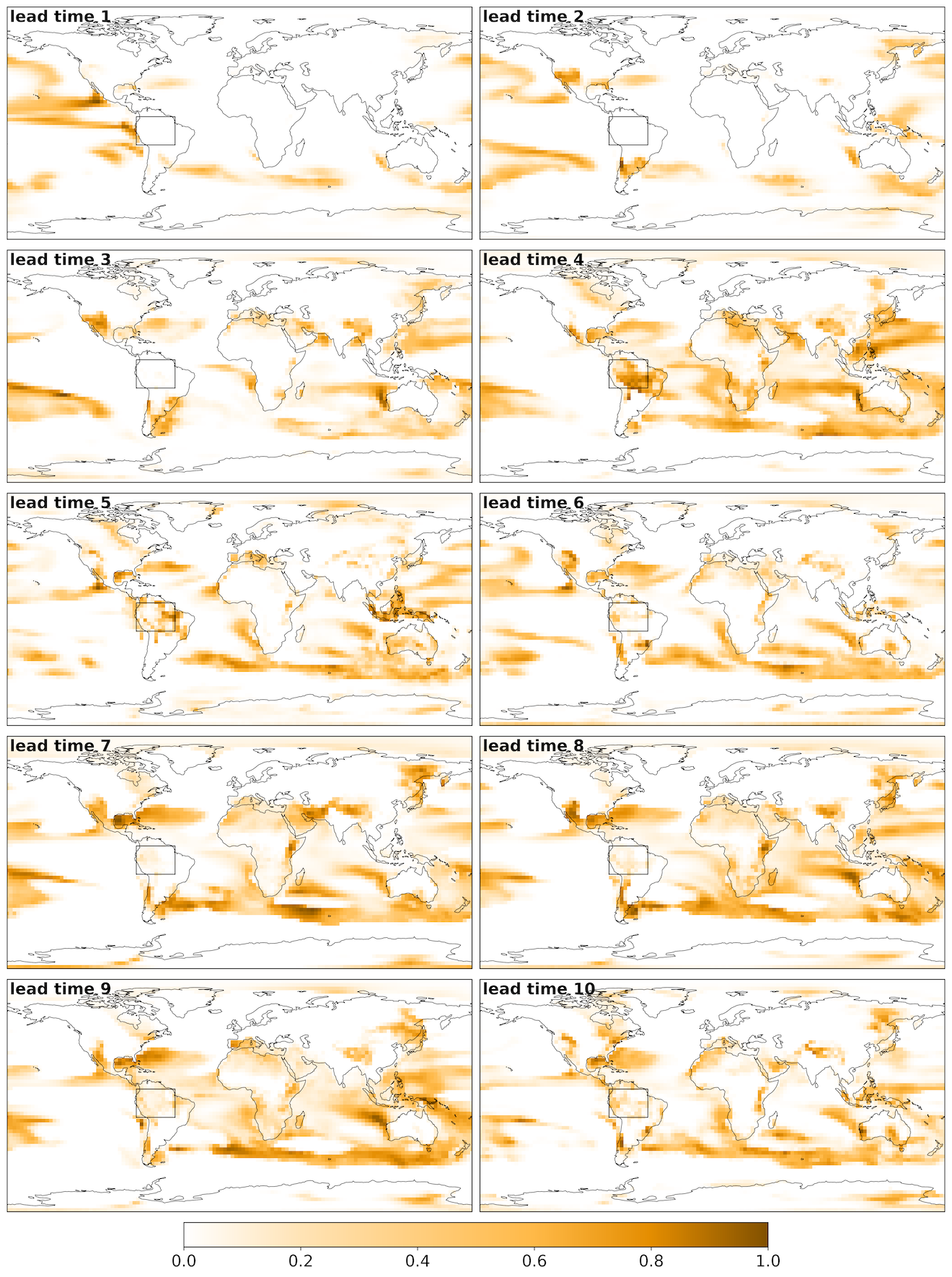}
    \caption[Amazon transfer learned masks, 1-10 years]{
        Transfer learned masks for the Amazon region for lead times 1-10 years.
    }
    \label{fig:amzn_all_tl_masks}
\end{figure}

\begin{figure}[!ht]
    \centering
    \noindent\includegraphics[width=0.85\textwidth]{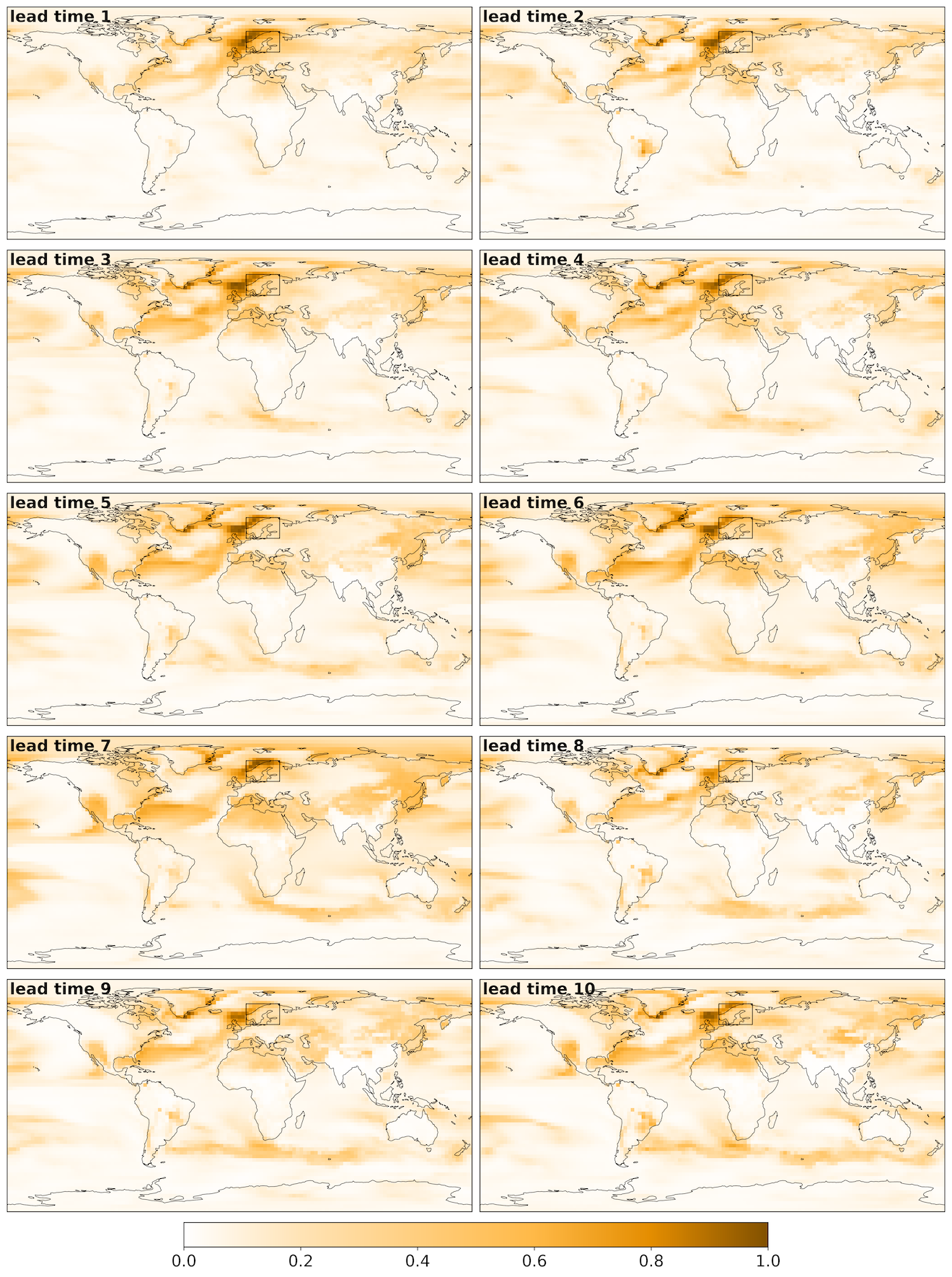}
    \caption[Northern Europe learned masks, 1-10 years]{
        Model learned masks for the northern Europe region for lead times 1-10 years.
    }
    \label{fig:neur_all_learned_masks}
\end{figure}

\begin{figure}[!ht]
    \centering
    \noindent\includegraphics[width=0.85\textwidth]{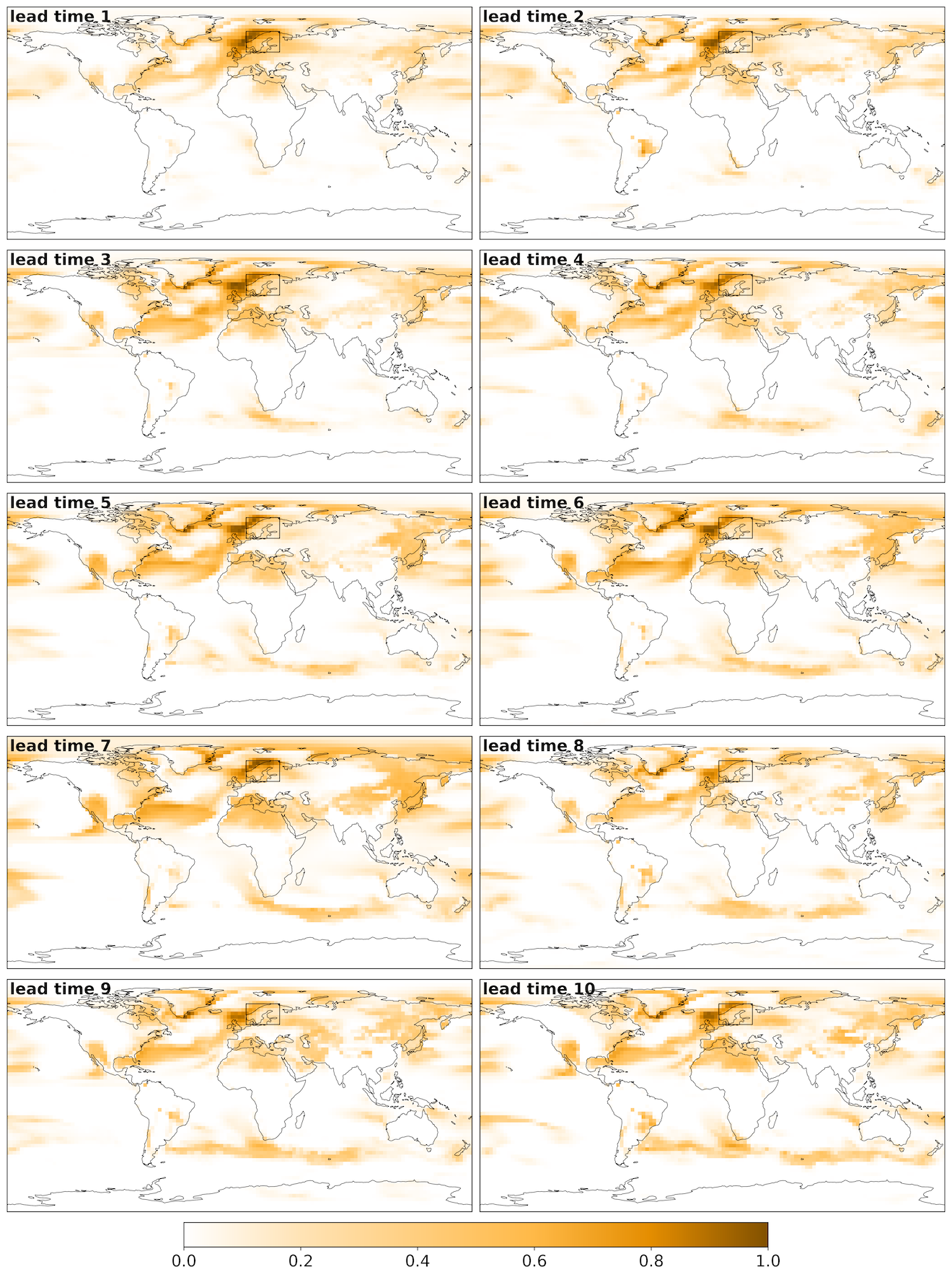}
    \caption[Northern Europe transfer learned masks, 1-10 years]{
        Transfer learned masks for the northern Europe region for lead times 1-10 years.
    }
    \label{fig:neur_all_tl_masks}
\end{figure}

\begin{figure}[!ht]
    \centering
    \noindent\includegraphics[width=0.85\textwidth]{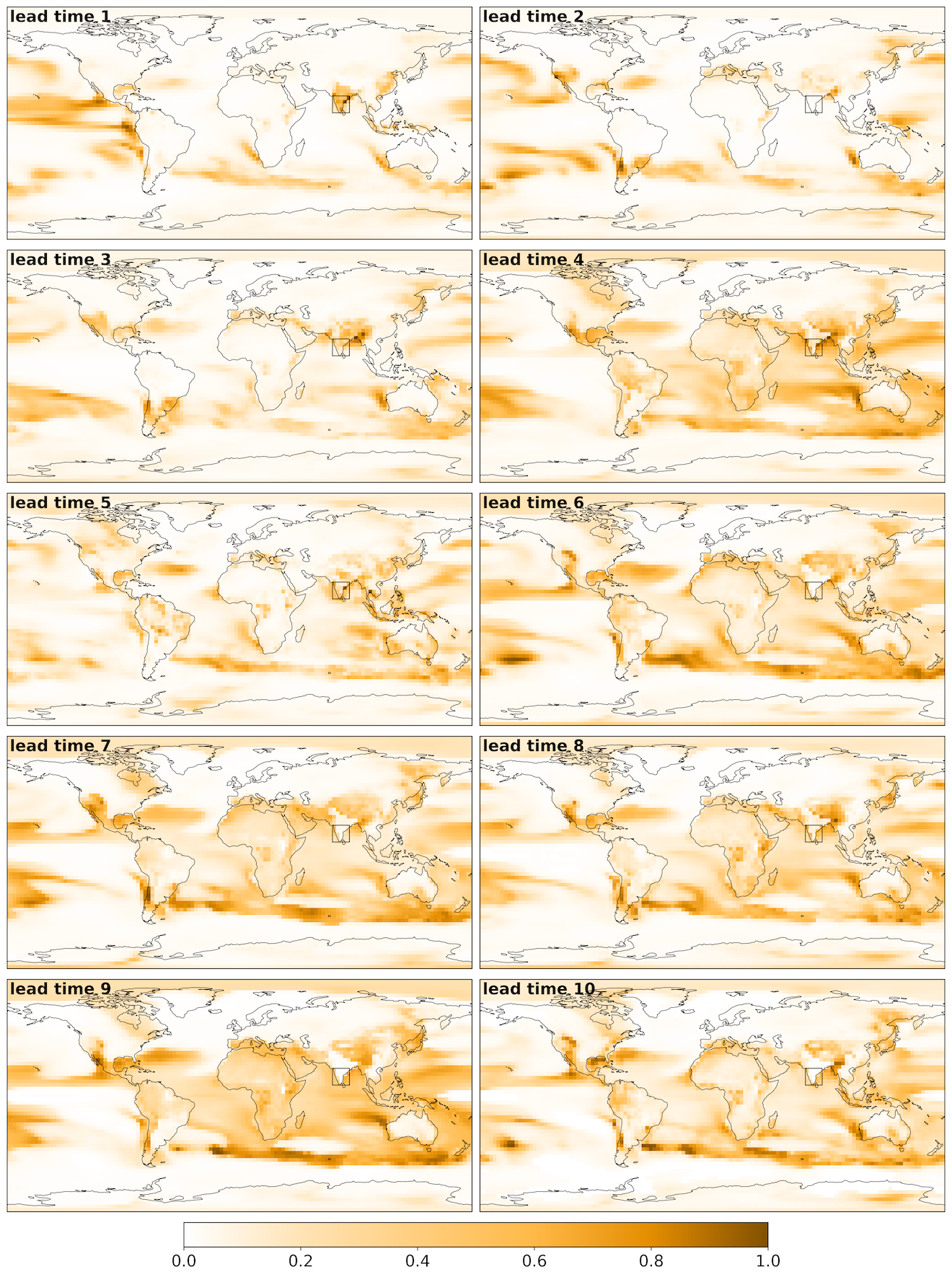}
    \caption[Southern India learned masks, 1-10 years]{
        Model learned masks for the southern India region for lead times 1-10 years.
    }
    \label{fig:sind_all_learned_masks}
\end{figure}

\begin{figure}[!ht]
    \centering
    \noindent\includegraphics[width=0.85\textwidth]{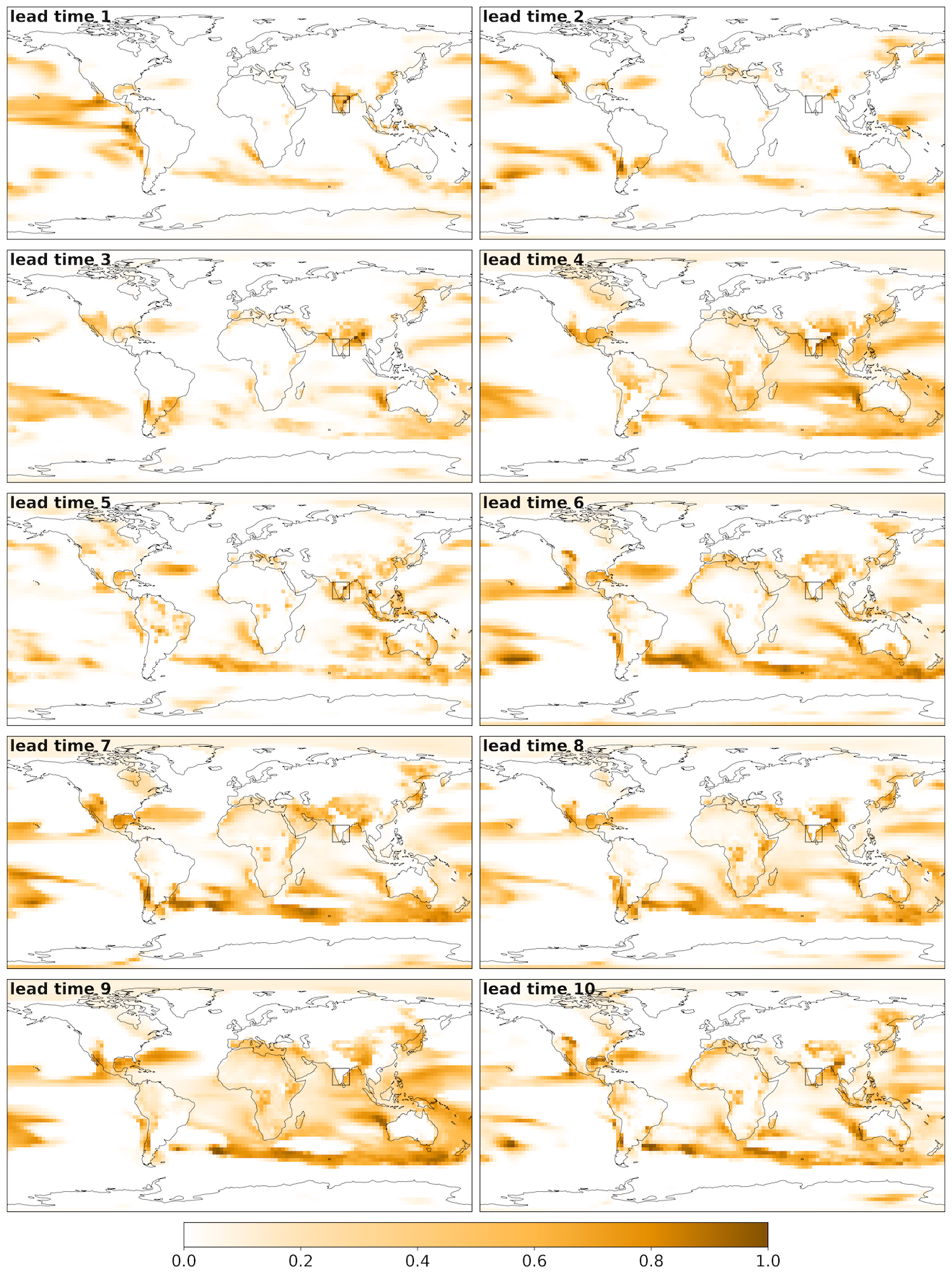}
    \caption[Southern India transfer learned masks, 1-10 years]{
        Transfer learned masks for the southern India region for lead times 1-10 years.
    }
    \label{fig:sind_all_tl_masks}
\end{figure}

\begin{figure}[!ht]
    \centering
    \noindent\includegraphics[width=0.85\textwidth]{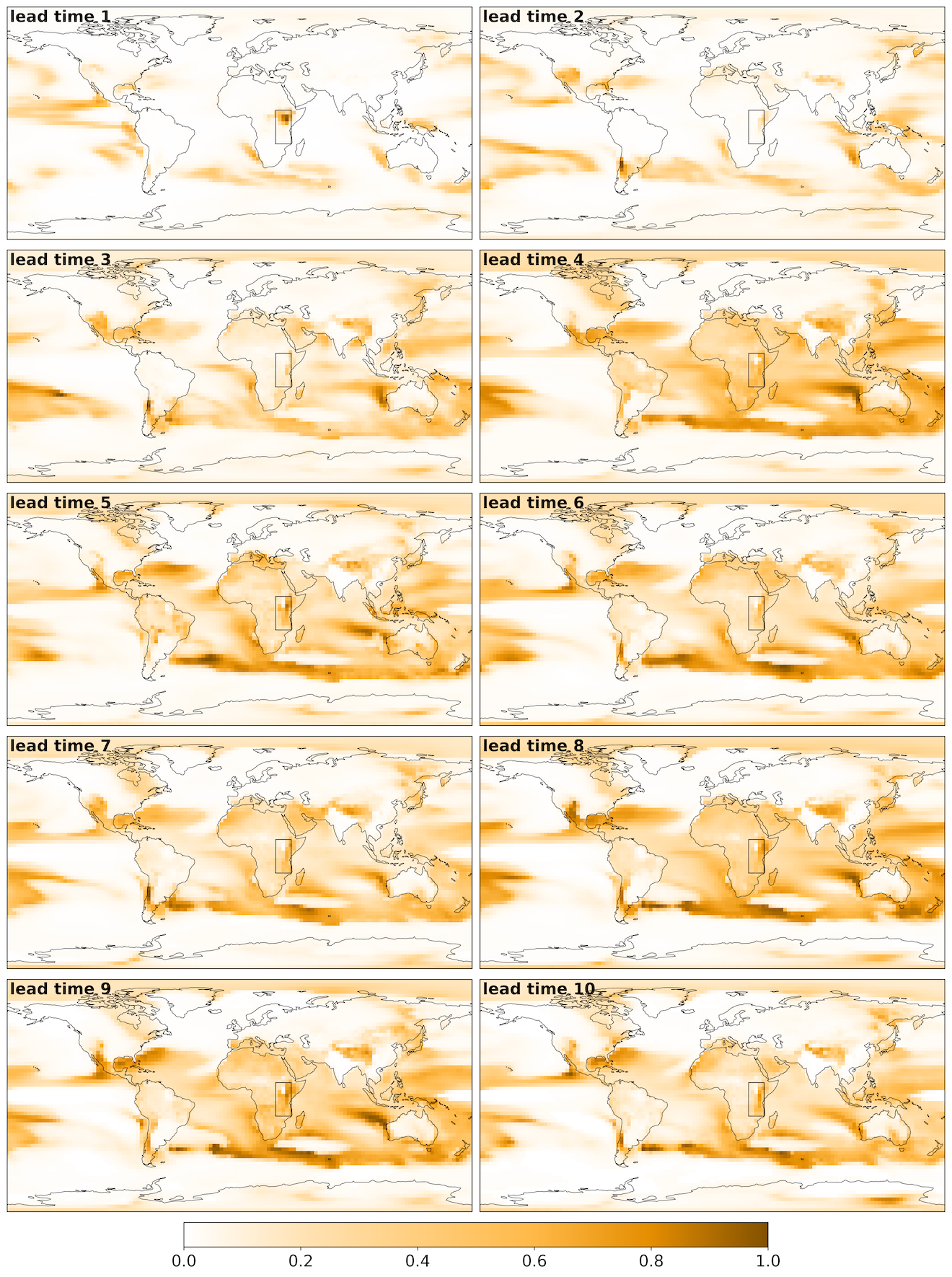}
    \caption[African Great Lakes learned masks, 1-10 years]{
        Model learned masks for the African Great Lakes region for lead times 1-10 years.
    }
    \label{fig:glaf_all_learned_masks}
\end{figure}

\begin{figure}[!ht]
    \centering
    \noindent\includegraphics[width=0.85\textwidth]{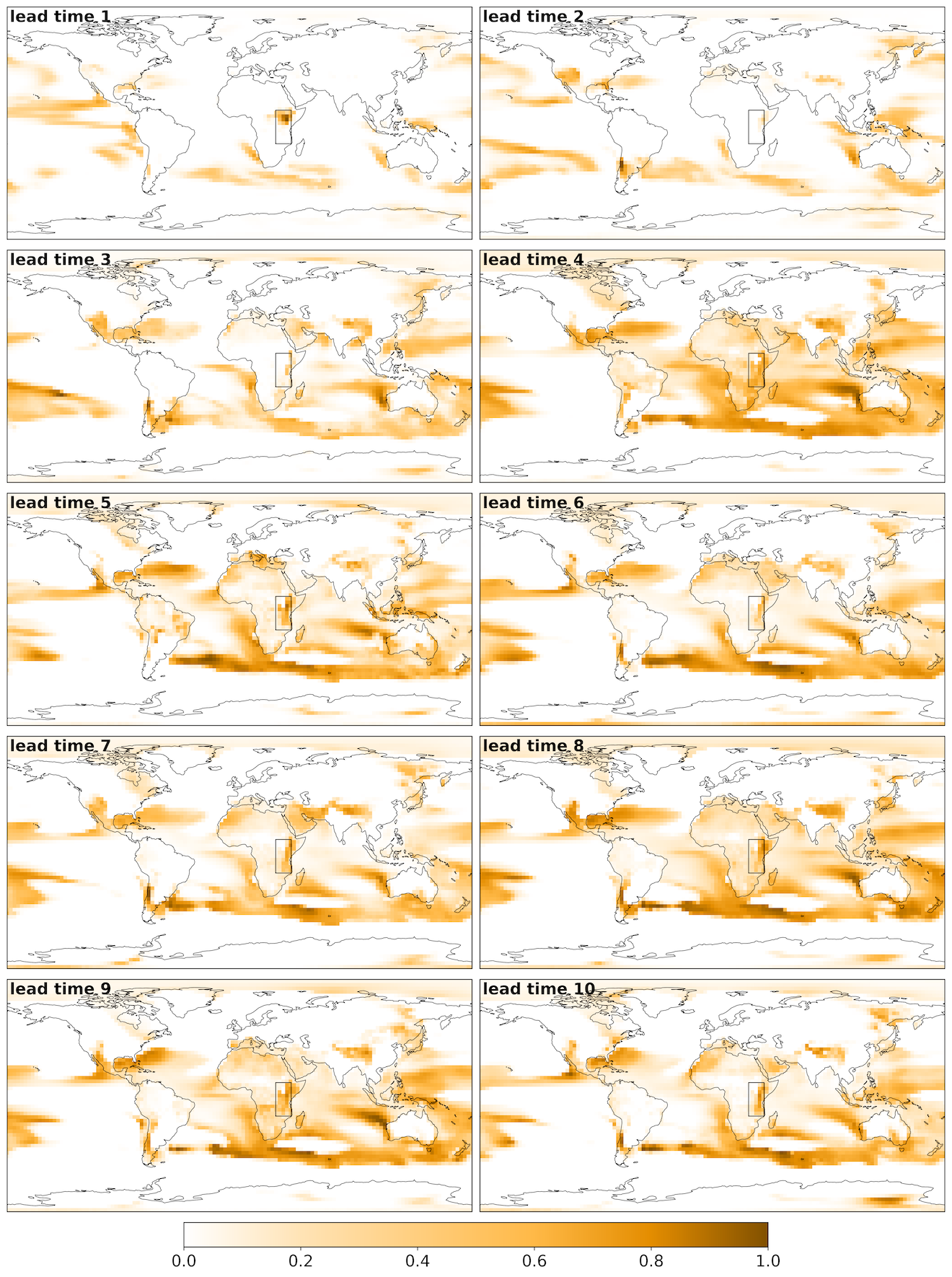}
    \caption[African Great Lakes transfer learned masks, 1-10 years]{
        Transfer learned masks for the African Great Lakes region for lead times 1-10 years.
    }
    \label{fig:glaf_all_tl_masks}
\end{figure}
\clearpage


\section{Transfer Learning}\label{sec:transfer learning}

\begin{figure}[!ht]
    \centering
    \noindent\includegraphics[width=0.47\textwidth]{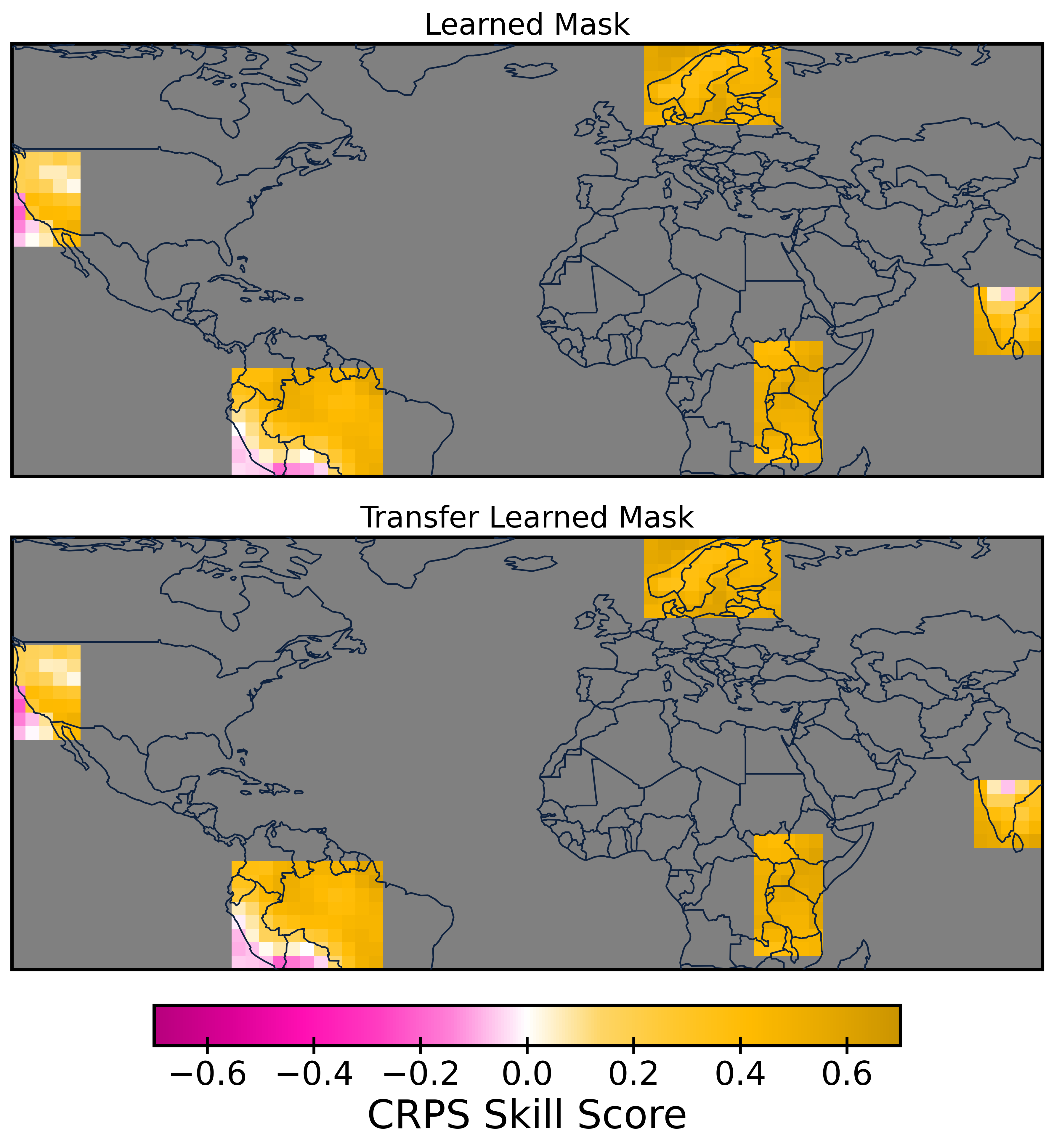}
    \noindent\includegraphics[width=0.47\textwidth]{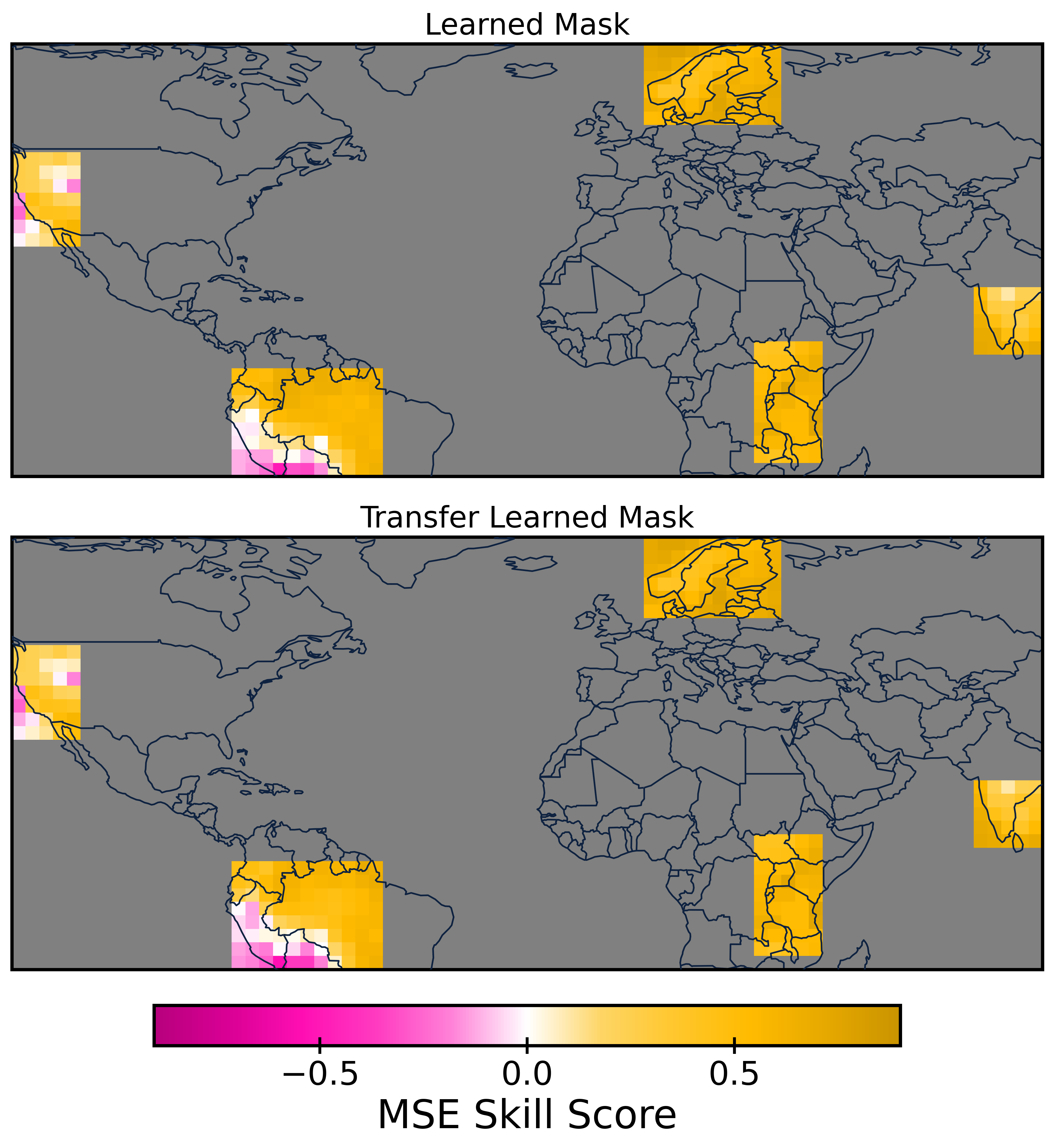}
    \caption[Transfer learned mask skill comparison]{
        CRPS (left) and MSE (right) skills score maps for the learned mask (top) and transfer learned mask (bottom).
        Despite the change in the masks (see the figures in Section \ref{sec:masks}), there is little to no change in the quality of the predictions for all regions.
    }
    \label{fig:transfer_learning}
\end{figure}
\clearpage


\section{Raw Metrics \& Bootstrapping}\label{sec:bootstrapping}

\begin{figure}[!ht]
    \centering
    \noindent\includegraphics[width=0.9\textwidth]{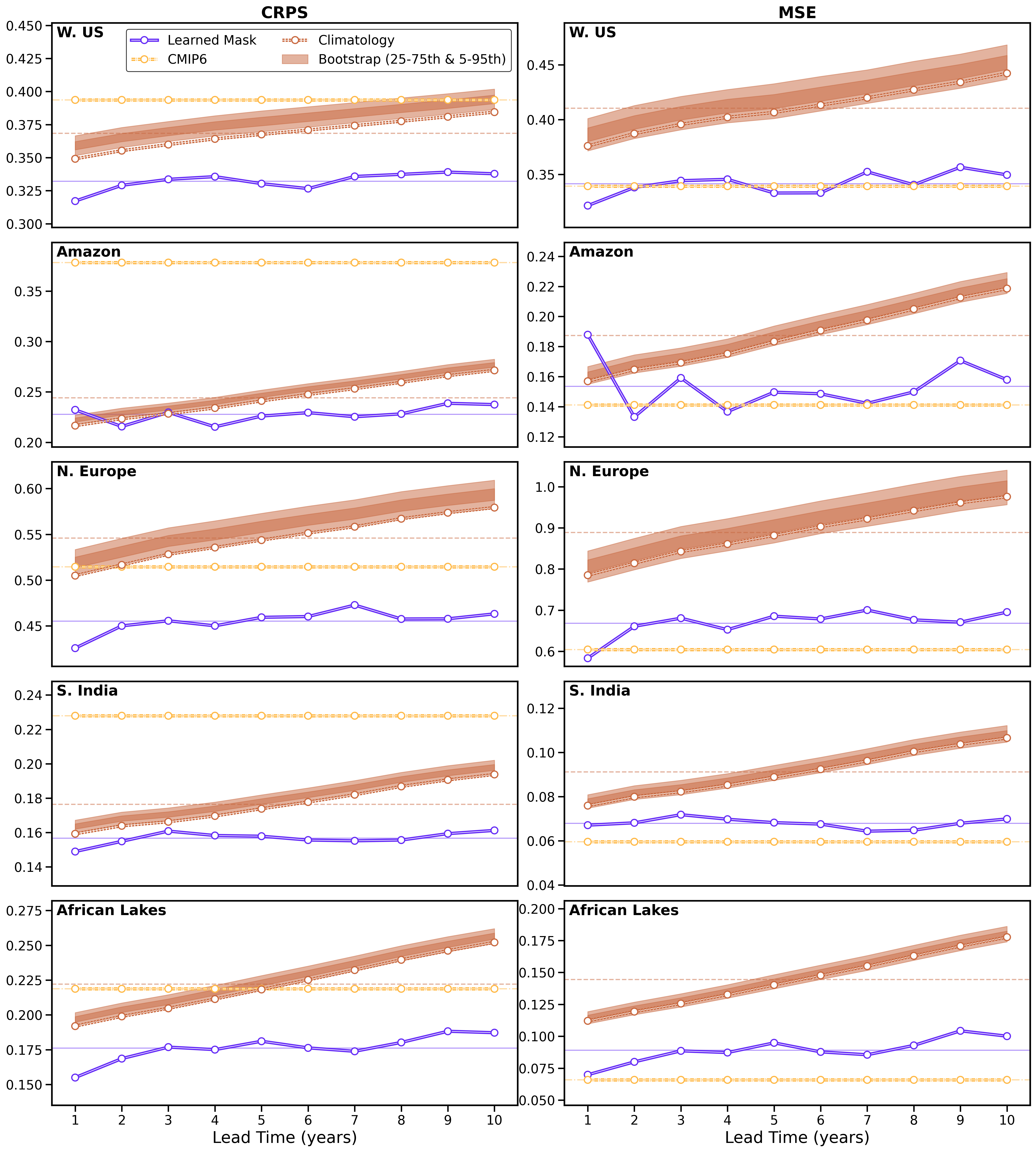}
    \caption[Raw metrics and bootstrapping results for five regions, 1931-2023]{
        Raw metrics for the reference climatology, learned mask analogs, and CMIP6 library.
        Also shown is the climatology bootstrapping, with shading indicating the 5th-95th percentile (light) and 25th-75th percentile (dark).
    }
    \label{fig:regions_sig_full}
\end{figure}

We show results as skill scores throughout the main manuscript, however, in order to explain the trend of increasing skill with lead time seen in many results, we plot the raw metrics in Figure \ref{fig:regions_sig_full}.
While the CMIP6 predictions are constant for all lead times (the same predictions are available for all dates, regardless of lead time), the climatology gets worse with lead time.
This leads to the trend of CMIP6 skill increasing with lead time (and to a lesser extent, the analog methods), as seen in the main manuscript results.

We also show the results of bootstrapping the reference climatology.
The bootstrapping samples the 30-year window up to and including the initialization state with replacement $10,000$ times.
The 5th to 95th percentile (light shading) and 25th to 75th percentile (darker shading) of the resulting distributions for MSE and CRPS are shown in Figure \ref{fig:regions_sig_full}.
We consider predictions that are lower than the 5th percentile and the climatology to be robust, which is indicated in the main manuscript results.
Note, the climatology prediction is always low (better performing) in the bootstrap distribution, indicating that using each sample in the 30-year window once is a good choice.

\begin{figure}[!ht]
    \centering
    \noindent\includegraphics[width=0.9\textwidth]{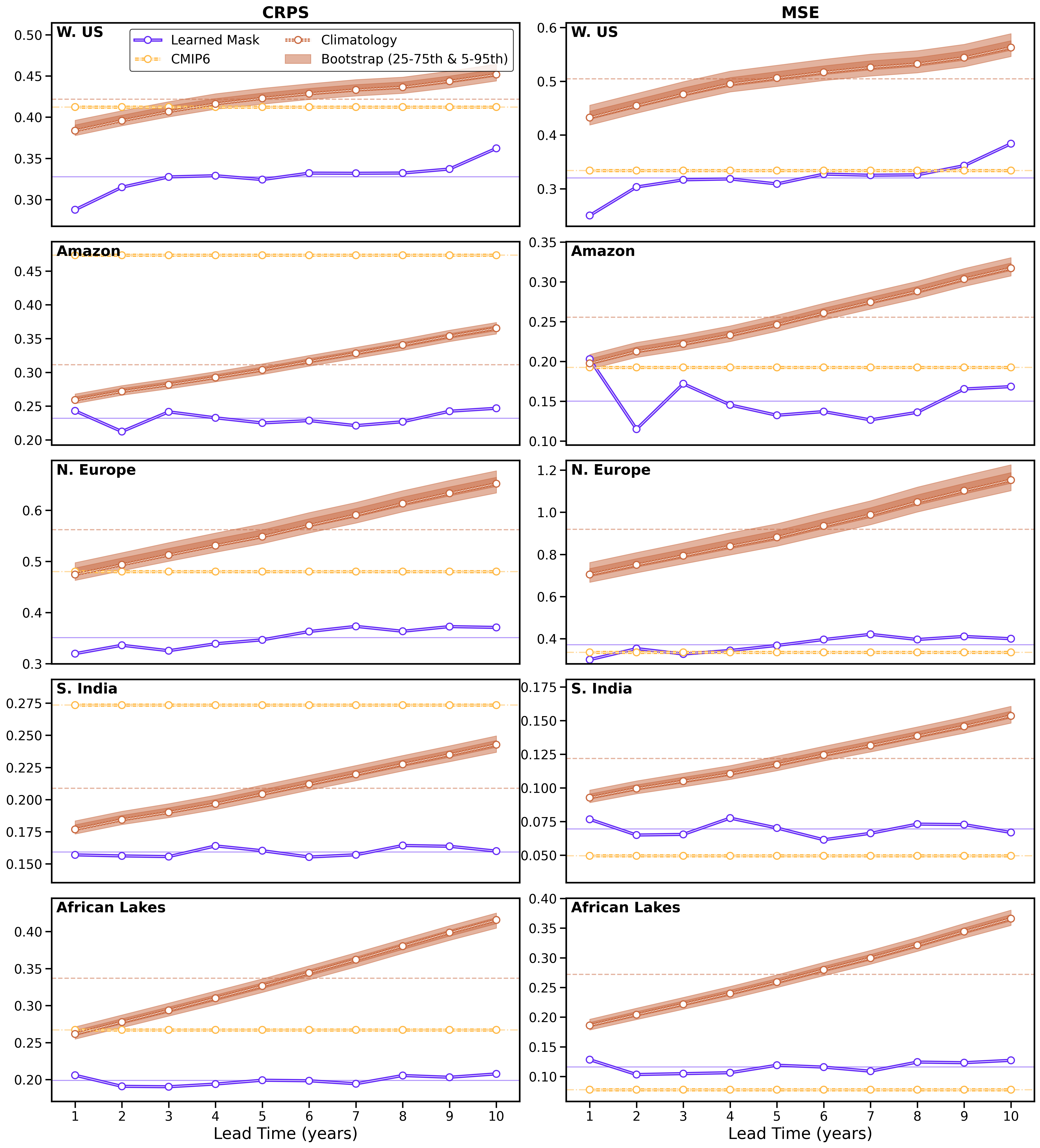}
    \caption[Raw metrics and bootstrapping results for five regions, 1994-2018]{
        Same as Figure \ref{fig:regions_sig_full}, but for 1994-2018.
    }
    \label{fig:regions_sig_iesm}
\end{figure}
\clearpage


\section{Spatial Metrics}\label{sec:spatial_metrics}

\begin{figure}[h!]
    \centering
    \noindent\includegraphics[width=0.85\textwidth]{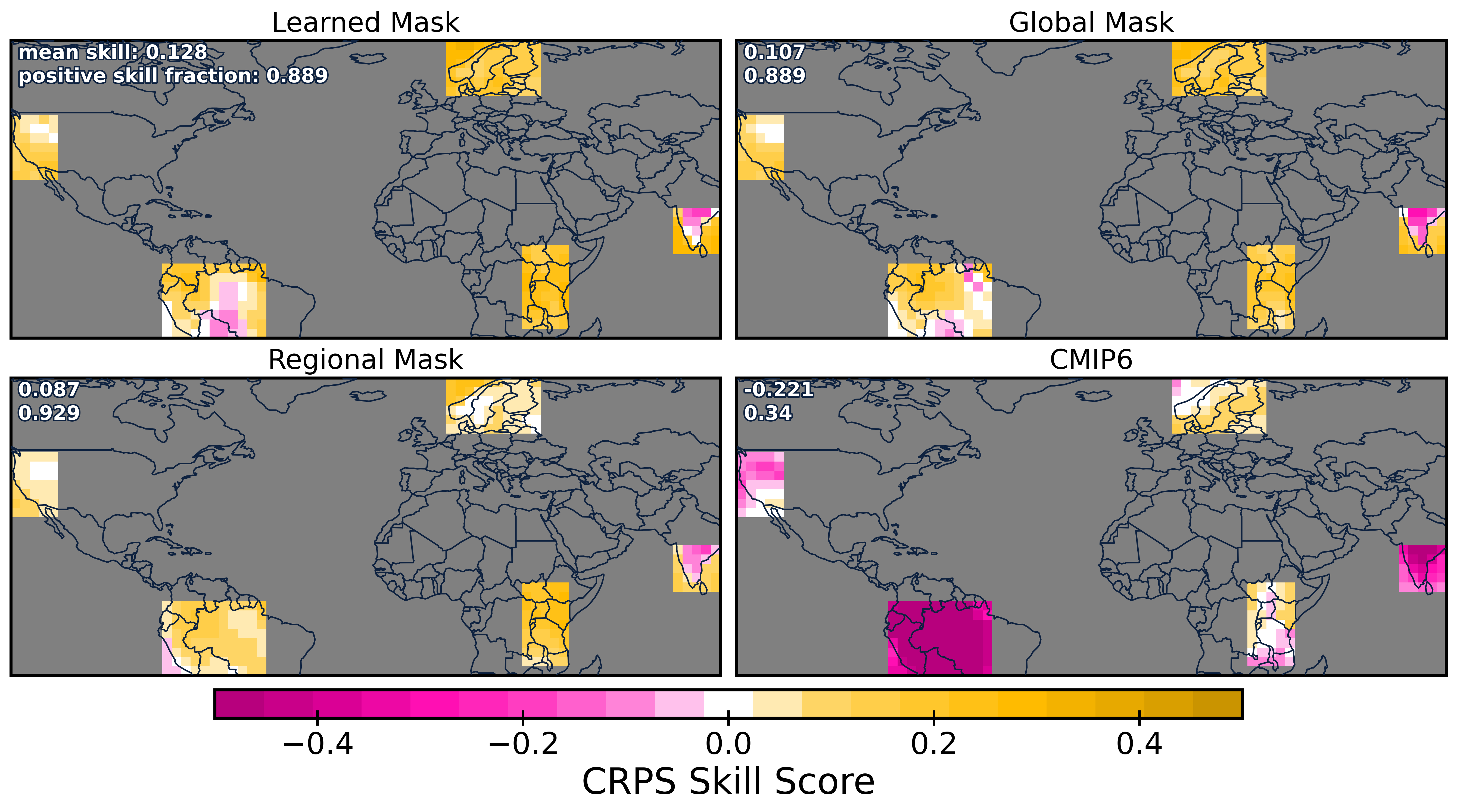}
    \\
    \noindent\includegraphics[width=0.85\textwidth]{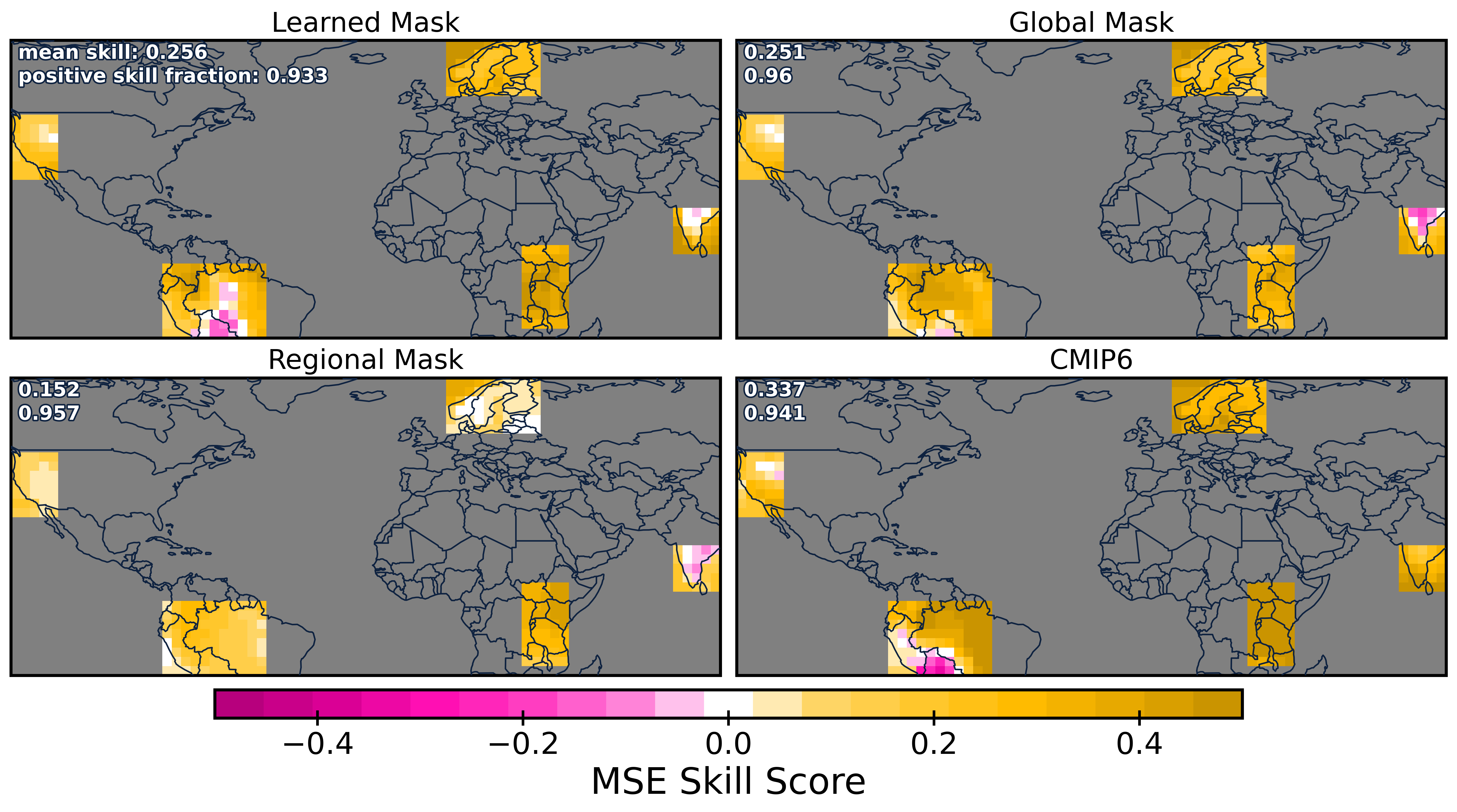}
    \caption[Spatial CRPS and MSE skill scores, 1931-2023]{Spatial CRPS (top) and MSE (bottom) skill scores.
    Skill is averaged over lead times 1-10 years and the time period 1931-2023 (93 years, selected such that all grid cells are available for climatology).
    The mean skill over the five regions and the fraction of grid cells with positive skill are reported for each method in the upper left of each panel.
    }
    \label{fig:full_region_skill}
\end{figure}

\end{document}